\documentclass[sigplan,screen]{acmart}

\usepackage[]{hyperref}
\usepackage{graphicx} 
\usepackage{subcaption}
\usepackage{float}
\usepackage{multicol}
\usepackage{threeparttable}
\usepackage{tabularx}
\usepackage{algorithmic}
\usepackage[ruled,vlined,linesnumbered]{algorithm2e}
\usepackage{subfloat}
\usepackage{enumitem}
\usepackage{placeins}
\usepackage{balance}

\copyrightyear{2026}
\acmYear{2026}
\setcopyright{acmlicensed}\acmConference[ASPLOS '26]{Proceedings of the 31st ACM International Conference on Architectural Support for Programming Languages and Operating Systems, Volume 1}{March 22--26, 2026}{Pittsburgh, PA, USA}
\acmBooktitle{Proceedings of the 31st ACM International Conference on Architectural Support for Programming Languages and Operating Systems, Volume 1 (ASPLOS '26), March 22--26, 2026, Pittsburgh, PA, USA}

\begin{document}

\title{
 GFS: A Preemption-aware Scheduling Framework for GPU Clusters with Predictive Spot Instance Management
}

\author{Jiaang Duan}
\authornote{Both authors were interns at Alibaba Group.}
\affiliation{%
  \institution{Shanghai Jiao Tong University}
  \city{Shanghai}
  \country{China}
}

\author{Shenglin Xu}
\authornotemark[1]
\affiliation{%
  \institution{Shanghai Jiao Tong University}
  \city{Shanghai}
  \country{China}
}

\author{Shiyou Qian}
\authornote{Corresponding authors: qshiyou@sjtu.edu.cn, yangdingyu@zju.edu.cn}
\affiliation{%
  \institution{Shanghai Jiao Tong University}
  \city{Shanghai}
  \country{China}
}

\author{Dingyu Yang}
\authornotemark[2]
\affiliation{%
  \institution{The State Key Laboratory of Blockchain and Data Security, Zhejiang University}
  \city{Hangzhou}
  \country{China}
}

\author{Kangjin Wang}
\affiliation{%
  \institution{Alibaba Group}
  \city{Hangzhou}
  \country{China}
}

\author{Chenzhi Liao}
\affiliation{%
  \institution{Alibaba Group}
  \city{Hangzhou}
  \country{China}
}

\author{Yinghao Yu}
\affiliation{%
  \institution{Alibaba Group}
  \city{Hangzhou}
  \country{China}
}

\author{Qin Hua}
\affiliation{%
  \institution{Shanghai Jiao Tong University}
  \city{Shanghai}
  \country{China}
}

\author{Hanwen Hu}
\affiliation{%
  \institution{Shanghai Jiao Tong University}
  \city{Shanghai}
  \country{China}
}

\author{Qi Wang}
\affiliation{%
  \institution{Alibaba Group}
  \city{Hangzhou}
  \country{China}
}

\author{Wenchao Wu}
\affiliation{%
  \institution{Alibaba Group}
  \city{Hangzhou}
  \country{China}
}

\author{Dongqing Bao}
\affiliation{%
  \institution{Alibaba Group}
  \city{Hangzhou}
  \country{China}
}

\author{Tianyu Lu}
\affiliation{%
  \institution{Alibaba Group}
  \city{Hangzhou}
  \country{China}
}

\author{Jian Cao}
\affiliation{%
  \institution{Shanghai Jiao Tong University}
  \city{Shanghai}
  \country{China}
}

\author{Guangtao Xue}
\affiliation{%
  \institution{Shanghai Jiao Tong University}
  \city{Shanghai}
  \country{China}
}

\author{Guodong Yang}
\affiliation{%
  \institution{Alibaba Group}
  \city{Hangzhou}
  \country{China}
}

\author{Liping Zhang}
\affiliation{%
  \institution{Alibaba Group}
  \city{Hangzhou}
  \country{China}
}

\author{Gang Chen}
\affiliation{%
  \institution{Zhejiang University}
  \city{Hangzhou}
  \country{China}
}


\settopmatter{printacmref=true}

\renewcommand{\shortauthors}{Jiaang Duan et al.}

\begin{abstract}

The surge in large language models (LLMs) has fundamentally reshaped the landscape of GPU usage patterns, creating an urgent need for more efficient management strategies. While cloud providers employ spot instances to reduce costs for low-priority (LP) tasks, existing schedulers still grapple with high eviction rates and lengthy queuing times. To address these limitations, we present GFS, a novel preemptive scheduling framework that enhances service-level objective (SLO) compliance for high-priority (HP) tasks while minimizing preemptions to LP tasks. Firstly, GFS utilizes a lightweight forecasting model that predicts GPU demand among different tenants, enabling proactive resource management. Secondly, GFS employs a dynamic allocation mechanism to adjust the spot quota for LP tasks with guaranteed durations. Lastly, GFS incorporates a preemptive scheduling policy that prioritizes HP tasks while minimizing the impact on LP tasks. We demonstrate the effectiveness of GFS through both real-world implementation and simulations. The results show that GFS reduces eviction rates by 33.0\%, and cuts queuing delays by 44.1\% for LP tasks. Furthermore, GFS enhances the GPU allocation rate by up to 22.8\% in real production clusters. In a production cluster of more than 10,000 GPUs, GFS yields roughly \$459,715 in monthly benefits.

\end{abstract}

\ccsdesc[500]{Computer systems organization~Cloud computing}

\keywords{Spot instance; forecasting model; preemptive scheduling}

\maketitle 

\section{Introduction}

LLMs, such as GPT \cite{chatgpt} and Llama \cite{Llama-3}, have created an unprecedented demand for graphics processing unit (GPU) resources. As organizations increasingly rely on GPUs to train and deploy LLMs, efficient resource management has become critical challenges for cloud providers. Task types in GPU resource allocation are highly dependent on the specific business use cases and their corresponding resource needs. Typical GPU cluster schedulers prioritize high-priority (HP)\footnote{High priority primarily reflects the non-preemptive nature of resource allocation, including LLM training tasks and online inference tasks.} tasks with strict SLOs, often resulting in underutilized GPU capacity that could be allocated to low-priority (LP) tasks, e.g., batch tasks and experimental training.
Although cloud providers (e.g., AWS EC2 \cite{EC2_Spot}) have extended the spot instance model—originally designed for CPU-based workloads—to GPU resources. GPU spot instances refer to preemptible GPU resources offered at significant cost discounts (typically 60-90\% lower than on-demand pricing \cite{miao2024spotserve}). However, GPU spot instances in multi-tenant clusters face additional challenges due to the non-preemptive nature of HP tasks and GPU memory fragmentation.

The coexistence of HP and spot tasks complicates resource allocation. Existing schedulers struggle to balance conflicting objectives: ensuring SLOs for HP tasks, minimizing eviction rates and queuing delays for spot tasks, and maximizing overall cluster usage. These challenges are intensified by dynamic resource demands and the inherent tension in resource sharing. Additionally, spot tasks can be preempted at any time when HP tasks require resources. Although cloud platforms typically offer a grace period (e.g., 30 seconds) upon preemption, the likelihood of termination remains high, particularly during periods of increased demand for GPUs.

By evaluating the one-month trace data from a 10,365-GPU production cluster, we obtained three key insights (\S \ref{sec:back_observe}). Firstly, the eviction rate for spot tasks is high during peak demand periods. Importantly, when spot tasks are preempted before their guaranteed durations, cloud providers cannot charge for the costs, and task states cannot be saved due to the absence of a checkpoint. Secondly, the weak SLO guarantees of spot instances, coupled with first-fit scheduling heuristics\footnote{First-fit scheduling is a resource allocation strategy, which assigns the first available resource (such as a GPU) that meets the requirements to the task.}\cite{knauth2012spot}, results in suboptimal GPU allocation issues. Lastly, gang-scheduled\footnote{Gang scheduling is common in LLM tasks, where related threads or processes simultaneously run on different processors or pods.} LLM workloads face longer queue times than conventional deep learning workloads. 

Various studies have delved into enhancing GPU efficiency and ensuring reliable task performance, which can be primarily split into two categories.
First, several strategies focus on minimizing preemption overhead through runtime optimization \cite{Lightweight, PipeSwitch}. For instance, PipeSwitch \cite{PipeSwitch} employs lightweight context-switching mechanisms to accelerate task migration during preemption. 
Second, emerging proactive frameworks seek to balance efficiency and reliability through architectural innovations \cite{Gandiva, AntMan}. Recent advancements in GPU spatial partitioning \cite{FGD_atc23} enable the concurrent execution of HP and spot tasks with hardware-enforced isolation.
Nevertheless, these approaches, while effective in specific scenarios, often lead to suboptimal resource allocation and do not adequately address dynamic workload variations.

\begin{figure}[t]
    \centering
    \includegraphics[width=0.96\linewidth]{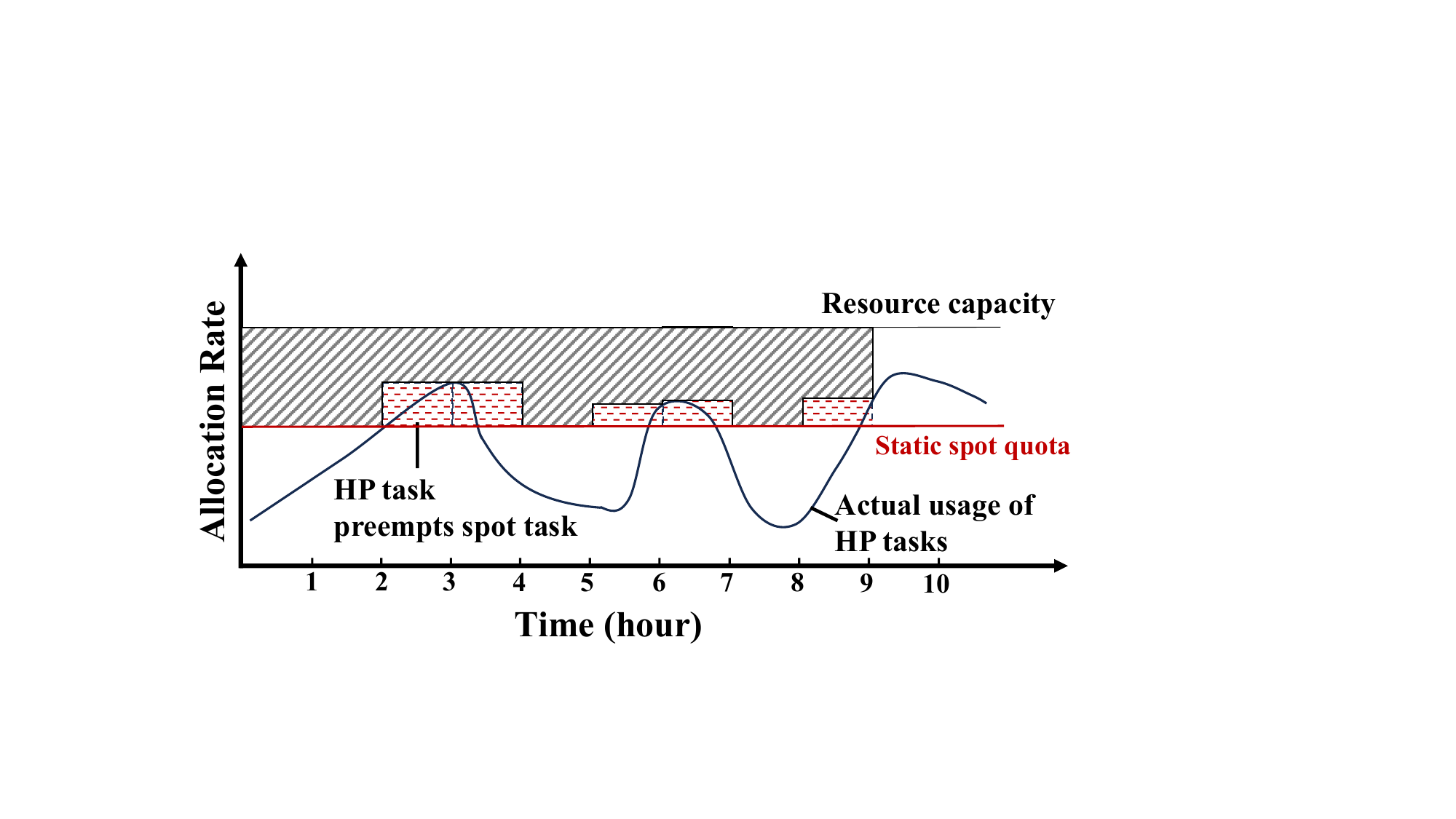}
    \caption{Diagram of spot evictions with a static spot quota.}
    \label{fig:spot_background}
\end{figure}

The key to enhancing GPU usage is innovative scheduling that addresses HP task SLOs and spot task eviction rates. However, three challenges require meticulous consideration. Firstly, effective resource allocation heavily relies on precise GPU demand forecasting, but existing models overlook organizational usage characteristics, and spatial-temporal correlations in cluster hotspots, critical factors impacting inventory estimation.
Secondly, current schedulers employ inflexible spot quota allocation that fails to adapt dynamically to real-time eviction rates and SLO compliance rates, especially during demand surges.
Thirdly, existing schedulers usually prioritize client-side cost optimizations while neglecting overall resource efficiency from the cloud providers' perspective.
Consequently, clusters frequently operate at suboptimal GPU allocations, leading to low overall resource usage and high spot eviction rates, as shown in Figure \ref{fig:spot_background}.

To address these issues, we present GFS (\S \ref{sec:design}), a proactive scheduling framework designed to enhance performance for both HP and spot tasks while maximizing cluster efficiency. 
Firstly, we develop a lightweight yet effective forecasting model that predicts GPU demand distributions by incorporating organizational patterns, temporal trends, and resource features. Secondly, we design a dynamic allocation mechanism that adjusts spot quotas based on anticipated GPU inventory, historical eviction rates, and queuing time. 
Finally, we propose a preemptive scheduling policy that considers preemption costs when allocating GPUs to HP tasks, effectively balancing the SLOs for both HP and spot tasks.

GFS has been implemented in production environments (\S \ref{sec:experiment}). Notably, for spot tasks, GFS achieves a 33.0\% reduction in eviction rates and a 44.1\% decrease in queuing times, while improving GPU allocation rates by 14.5\%.\footnote{GFS prioritizes the allocation rate of GPUs instead of their utilization.}
Additionally, we evaluate GFS through trace-driven simulations using real-world data. The results show that GFS reduces the average queuing time for HP tasks by $63.5\%$ and shortens the completion time for spot tasks by $14.5\%$, compared to four state-of-the-art schedulers. The source code for GFS is available at \url{https://github.com/Sjtucitlab/Spot}.

Our contributions can be summarized as follows:

\begin{itemize}[itemsep=2pt,topsep=0pt,parsep=0pt]
    \item We identify a dilemma in GPU  clusters: low overall GPU usage and frequent preemption of spot tasks. 

     \item We propose GFS, a proactive scheduling framework that efficiently allocates GPU resources across different task types and schedules them accordingly.

    \item We implement GFS in production GPU clusters and evaluate its performance through simulations, confirming the effectiveness of GFS. We have released a production GFS serving trace \footnote{\url{ https://github.com/alibaba/clusterdata/tree/master/}} as part of the Alibaba cluster trace program \cite{alibaba_clusterdata}.
\end{itemize}

The remainder of this paper is organized as follows. Section \ref{sec:back_observe} presents the background and observations. Section \ref{sec:design} describes the framework of GFS and its three modules. Section \ref{sec:experiment} analyzes the experimental results. Section \ref{sec:relatedwork} reviews related work, and Section \ref{sec:conclusion} concludes the work.

\section{Background and Observations}
\label{sec:back_observe}

\subsection{Background}

\textbf{GPU Resource Management and Spot Instances.}
Traditional GPU schedulers employ priority-based strategies to ensure SLOs for HP tasks, leading to resource underutilization. Studies have shown that average GPU utilization in production clusters rarely exceeds 30–40\%, primarily due to rigid resource reservations and conservative over-provisioning \cite{li2022ai}. Spot instances, inspired by cloud computing’s spot markets \cite{Bamboo}, aim to reclaim this idle capacity for LP tasks, subject to eviction when HP task demands arise. While this approach improves resource efficiency, existing implementations face critical limitations: frequent evictions, increased completion times, and violated soft SLOs \cite{Snape}. Moreover, reactive scheduling fails to handle demand surges, leading to suboptimal placement decisions and resource fragmentation \cite{DeepSpotCloud}.

\begin{figure}[b]
\vspace{-5mm}
    \centering
    \subfloat[CDF of pod GPU requests]{
    \includegraphics[width = 0.45\linewidth]{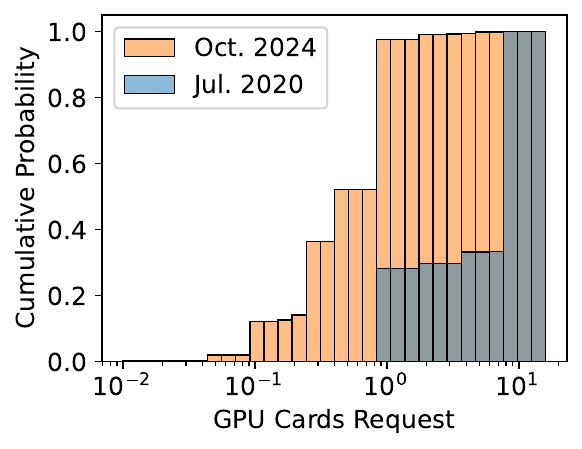}
    \label{fig:llm_inst_spec}
  }
  \hspace{2pt}
    \subfloat[CDF of task GPU requests]{
    \includegraphics[width=0.45 \linewidth]{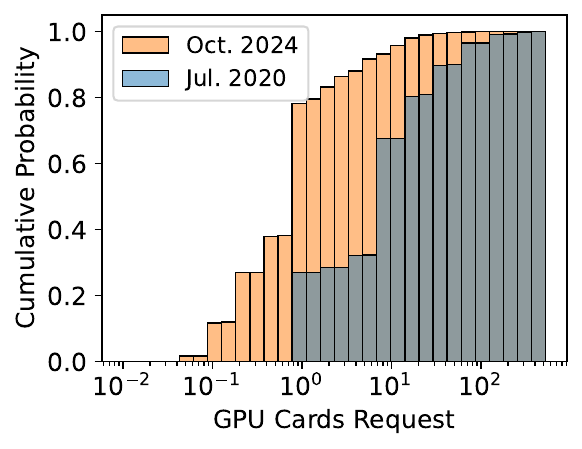}
     \label{fig:llm_task_spec}
    }
    \vspace{-3mm}
    \caption{CDF of GPU requests at the pod and task levels in Oct 2024 and Jul 2020.}
    \vspace{-2mm}
\end{figure}

\textbf{Demand Forecasting for Resource Allocation.}
Accurate GPU demand prediction is foundational to proactive resource management. Time-series forecasting models, such as ARIMA \cite{ARIMA}, Prophet \cite{Prophet}, and LSTM \cite{LSTM}, have been applied to predict cloud resource usage. However, these methods often overlook organizational hierarchies and cost-driven usage patterns inherent in production environments. For instance, GPU consumption in large organizations may vary significantly across departments due to budget cycles, project deadlines, or shifting business priorities, a dimension rarely captured by conventional models. Furthermore, most existing methods generate point estimates rather than probabilistic forecasts, limiting their utility in risk-aware scheduling systems that must account for demand uncertainty \cite{jhin2024addressing}.

\subsection{Observations}

\begin{table}[t]
  \begin{center}
    \caption{GPU statistics in a production cluster.}
    \label{tab:node_preview}
    \resizebox{\columnwidth}{!}{
    \begin{tabular}{cccc}
      \toprule
      \textbf{GPU Model} & \textbf{Node Number} & \textbf{GPUs/Node}  & \textbf{Allocation Rate}\\
      \midrule
      A10 & >2000 & 1 & \textbf{84.59\%}\\ 
      A100 & >400 & 8 & \textbf{74.34\%}\\ 
      A800 & >50 & 8 & \textbf{62.96\%}\\
      H800 & >200 & 8 & \textbf{68.11\%}\\
      \bottomrule
    \end{tabular}
    }
  \end{center}
  \vspace{-5mm}
\end{table}

\begin{figure*}[t!]
    \centering
    \begin{minipage}{0.6\linewidth}
        \centering
        \includegraphics[width=0.8\linewidth]{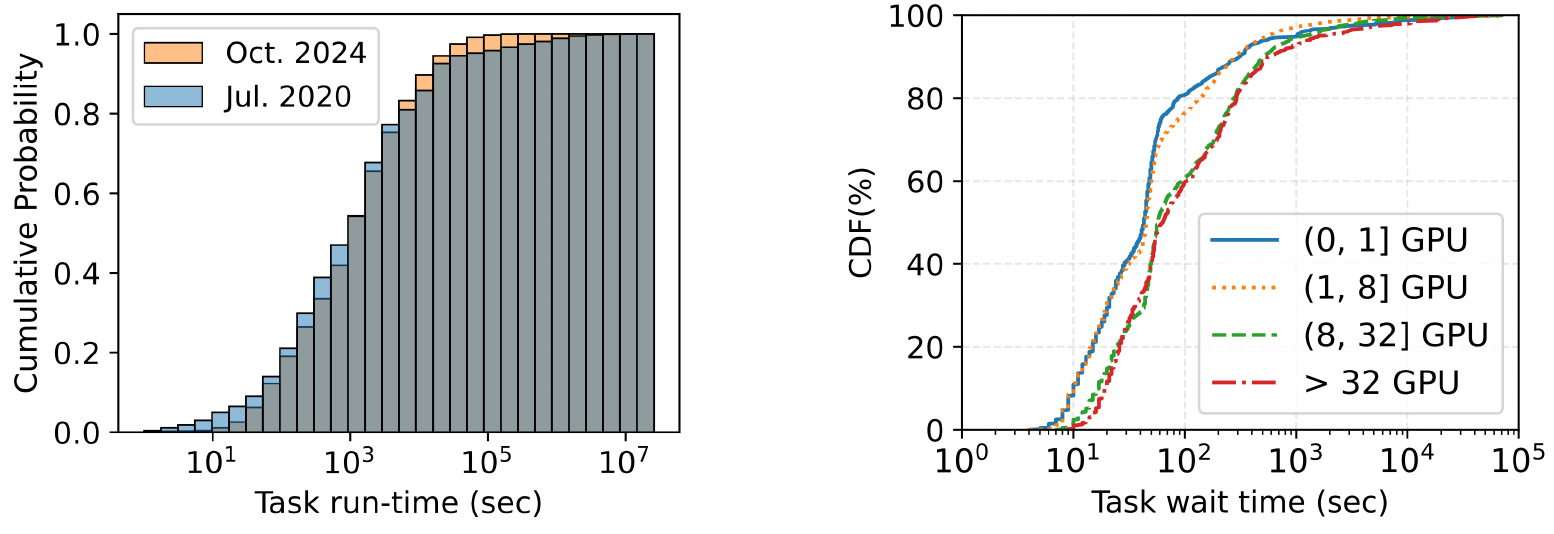}
        \vspace{-1mm}
        \caption{Running and queuing time of tasks in production GPU cluster.}
        \label{fig:job_running_time}
    \end{minipage}
    \hspace{4pt}
    \begin{minipage}{0.35\linewidth}
        \centering
        \includegraphics[width=0.77\linewidth]{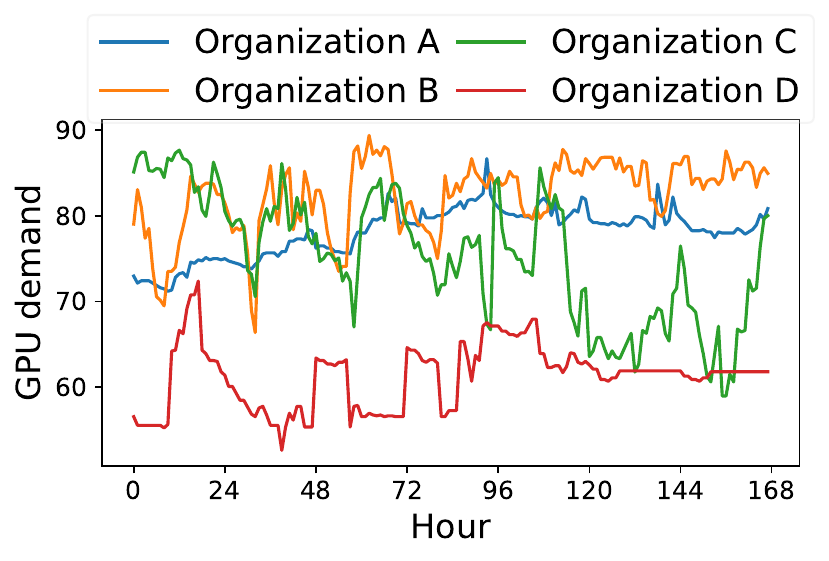}
        \vspace{-4mm}
        \caption{GPU request of four organizations.}
        \label{fig:demand}
    \end{minipage}
    \vspace{-4mm}
\end{figure*}

To investigate and address the challenges posed by the rapid rise of LLMs on HP and spot tasks, we analyzed the real data in Oct 2024 from a production cluster consisting of 10,365 GPUs, featuring heterogeneous GPU models as shown in Table \ref{tab:node_preview}. Our analysis reveals three insights.

\emph{\textbf{Observation $1$:} The emergence of LLMs has fundamentally changed GPU resource demand patterns.}

Figures \ref{fig:llm_inst_spec} and \ref{fig:llm_task_spec} compare the CDF distributions of requested GPU resources at the pod and task levels in Jul 2020 and Oct 2024 respectively in production clusters. Three systemic pressure points are identified. 
First, nearly 100\% of pods in 2024 require full GPU cards, compared to 80\% partial-card requests in 2020, with 70\% demanding full-node 8-GPU allocations, marking a 3.2× increase from 2020. Task-level GPU requests show a similar trend.
Second, LLM workloads exhibit extreme temporal persistence, with P90/P99 runtimes reaching 6.4 hours and 19.8 days, respectively, as shown in Figure~\ref{fig:job_running_time}, which are 1.44$\times$ and 21.17$\times$ longer than the 2020 baselines. This leads to "GPU hoarding" effects, reducing resource turnover and worsening scarcity. 
Third, gang-scheduled tasks face notable queuing times, as shown in Figure \ref{fig:job_running_time}, with 8-GPU tasks having a median waiting time of 4.8 hours, which is 2.7$\times$ longer than that of 1-GPU tasks.

\emph{\textbf{Observation $2$:}The weak SLO guarantees of spot tasks, coupled with first-fit scheduling heuristics, result in suboptimal GPU allocation.}

Table \ref{tab:node_preview} shows the GPU resource distribution and average allocation rates for more than 10000 GPUs in the cluster. Two significant systemic inefficiencies hinder SLO compliance and resource usage.  
First, the use of first-fit scheduling heuristics, along with weak SLO guarantees (e.g., high eviction rate) for spot tasks, leads to fragmented GPU allocation. This is evidenced from the low allocation rates of high-end GPUs, such as A100, A800, and H800, all of which fall below 80\%. 
These statistics indicate that a substantial portion of resources remains underutilized due to ineffective allocation strategies, highlighting inefficient GPU usage.

Second, there is a significant relationship between the inefficient GPU usage and demand fluctuations. Figure \ref{fig:demand} shows the GPU requests of 4 organizations over time. Notable differences in GPU demand exist across organizations. For instance, the GPU demand of Organization A is generally stable, but shows clear peaks at certain periods, leading to frequent fluctuations in resource needs. The maximum number of GPUs requested reaches 86, while the minimum is 74. In contrast, Organization B exhibits more pronounced fluctuations, where the maximum number of GPUs requested is 90 and the minimum is 67.

Current scheduling strategies, like first-fit, fail to adapt to fluctuations in GPU demand. Consequently, GPUs are inefficiently allocated and their computing power is underutilized. This misalignment in resource allocation exacerbates SLO violations, particularly during times of high demand.

\begin{figure}[b!]
\vspace{-7mm}
    \centering
    \subfloat[Week 1]{
    \includegraphics[width = 0.47\linewidth]{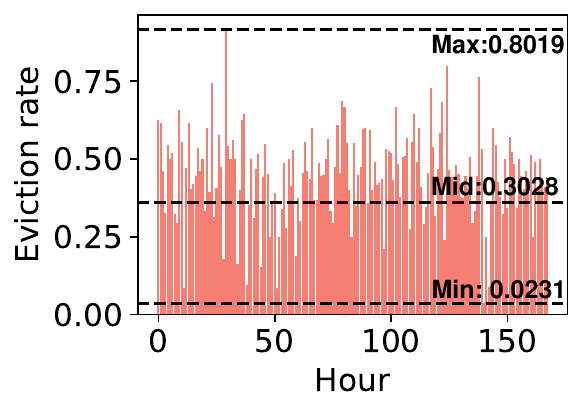}
    \label{fig:spot_evict_01}
  }
    \subfloat[Week 2]{
    \includegraphics[width = 0.47\linewidth]{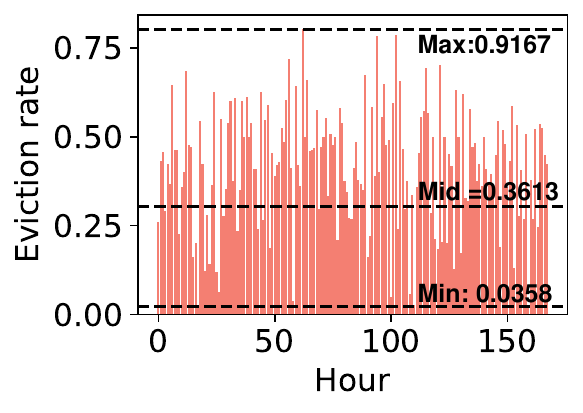}
    \label{fig:spot_evict_02}
  }
  
    \subfloat[Week 3]{
    \includegraphics[width=0.47\linewidth]{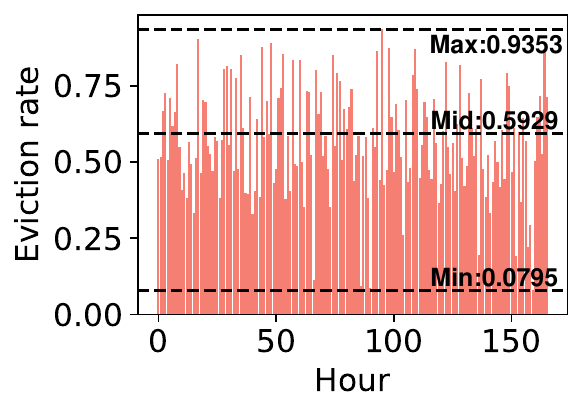}
     \label{fig:spot_evict_03}
    }
    \subfloat[Week 4]{
    \includegraphics[width=0.47\linewidth]{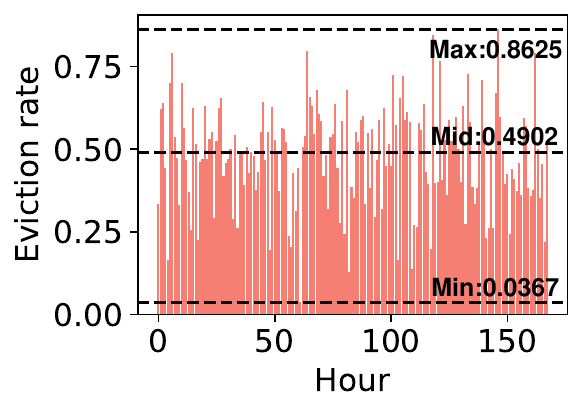}
     \label{fig:spot_evict_04}
    }
    \vspace{-3mm}
    \caption{Eviction rate of spot tasks in a production cluster over four weeks.}
    \label{fig:spot_job_eviction}
\end{figure}

\emph{\textbf{Observation $3$:} The eviction rate for spot tasks is high during peak demand periods.}

\label{para:preemption}

Figure \ref{fig:spot_job_eviction} illustrates the fluctuating eviction rates of spot tasks over four consecutive weeks in Oct 2024. Specifically, rates are higher from 10:00 to 12:00, peaking at 93.53\%, compared to much lower rates of 2.31\% during off-peak hours. On average, the eviction rate across all weeks is approximately 49.5\%, highlighting substantial instability.
Additionally, eviction rates vary across weeks. Week 3 exhibits the highest rate of 93.53\% and Week 1 records a lower rate of 80.19\%. 

These fluctuations can be attributed to two main factors. First, 68.3\% of evictions occur when HP tasks preemptively claim reserved quotas, suggesting inadequate resource allocation between HP and spot tasks. A strong positive correlation between eviction rates and HP task arrival intensity is observed, with a correlation coefficient of 0.87. Second, the absence of predictive HP task demand modeling for spot allocation results in a 41.2\% overcommitment of spot quotas during high-demand windows. These statistics underscore the dynamic nature of eviction rates and the need for proactive resource management.




\section{Design of GFS}
\label{sec:design}

Under heavy, dynamic, and diverse workloads, achieving efficient task scheduling is challenging. 
Current production GPU clusters often encounter a dilemma: tasks experience significant waiting time while the GPU allocation rate stays low. 
Firstly, there is currently no accurate algorithm for predicting GPU demand for HP tasks, nor an effective mechanism for managing spot quota. This leads to resource conflicts between spot and HP tasks. Secondly, when scheduling HP and spot pods in the cluster, these two types of pods do not avoid each other effectively, causing scheduling conflicts on preferred nodes. Thirdly, when spot tasks are preempted by HP tasks, there is no appropriate preemption strategy in place.
To address these issues, three key points must be considered. First, the GPU demands for HP tasks must be accurately predicted. Second, the quota for spot tasks should be dynamically adjusted alongside HP tasks. Thirdly, when necessary, spot tasks should be preempted at minimal costs. 

\begin{figure}[tb]
    \centering
    \includegraphics[width=\linewidth]{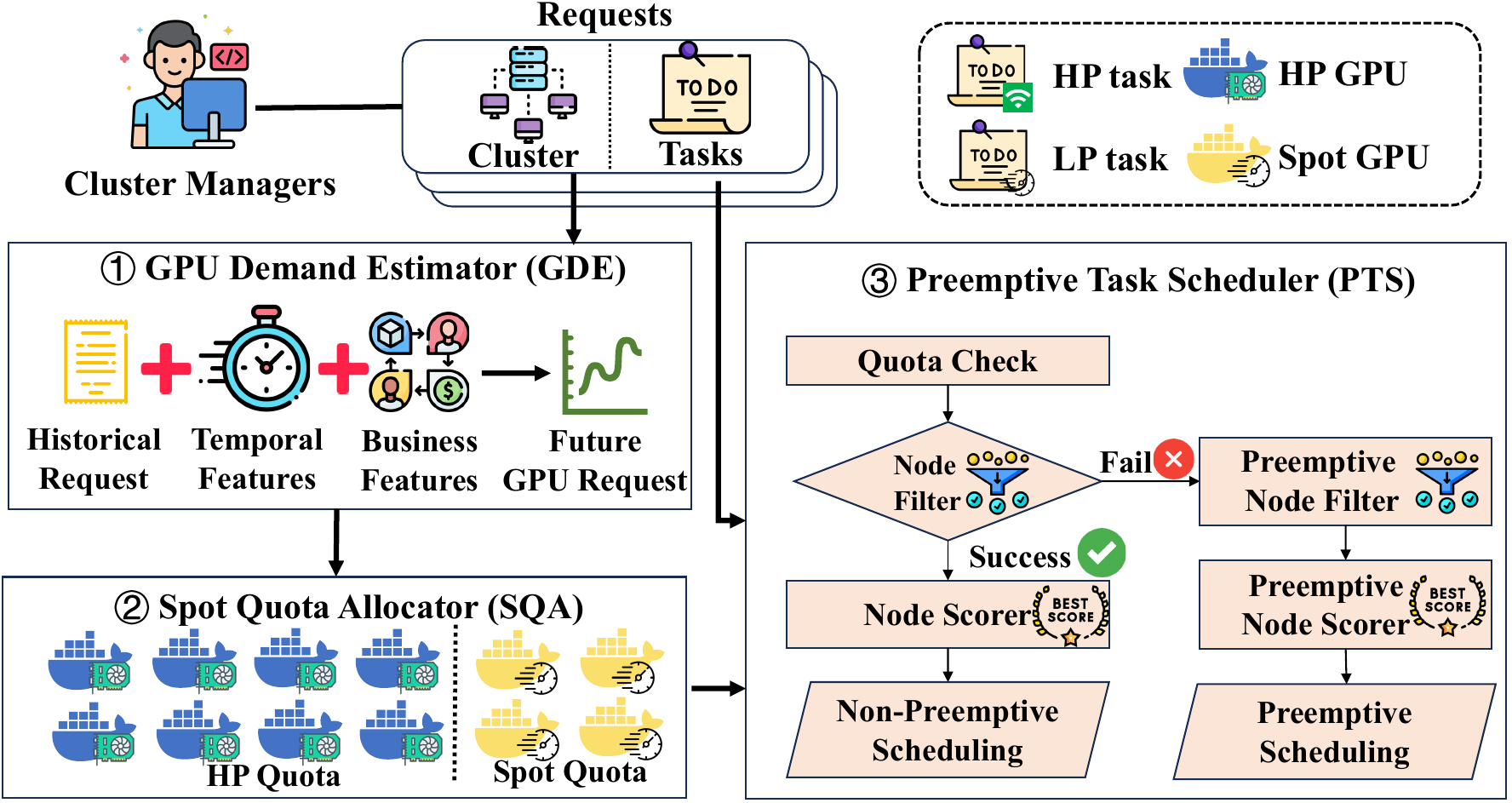}
    \vspace{-6mm}
    \caption{Framework of GFS}
    \label{fig:system_framework}
    \vspace{-6mm}
\end{figure}

\subsection{Overview}

The architecture of GFS, shown in Figure \ref{fig:system_framework}, comprises three main components: a GPU demand estimator (GDE), a spot quota allocator (SQA), and a preemptive task scheduler (PTS). 
First, by analyzing GPU usage patterns, temporal features, business metadata, and differentiated demand modeling across various business units, GDE employs an organization decomposed time-series prediction approach to model the probabilistic distribution of future GPU demand for HP tasks. 
Second, SQA dynamically adjusts GPU quotas for spot tasks through an eviction-aware feedback mechanism. It integrates GPU demand predictions, real-time cluster conditions, and spot SLOs to determine optimal quota limits. 
Third, PTS implements an efficient scheduling strategy that prioritizes non-preemptive allocation while allowing for controlled preemption.

The three components of GFS work together to enable proactive resource management and effective scheduling via a closed-loop design. 
For each new task (either HP or spot), SQA verifies availability of allocated quota. For HP tasks, PTS filters nodes with sufficient idle GPUs that meet the task requirements (e.g., GPU models and memory). Candidate nodes are scored from multiple dimensions, and the node with the highest score is chosen for task deployment. If no idle resources are available, PTS calculates the minimal preemption cost to identify nodes running spot tasks that can be preempted to free resources. For spot tasks, PTS locates nodes with available spot GPUs.

\subsection{GPU Demand Estimator (GDE)}

\label{sec:des}

\begin{figure}[t]
    \centering
    \includegraphics[width=\linewidth]{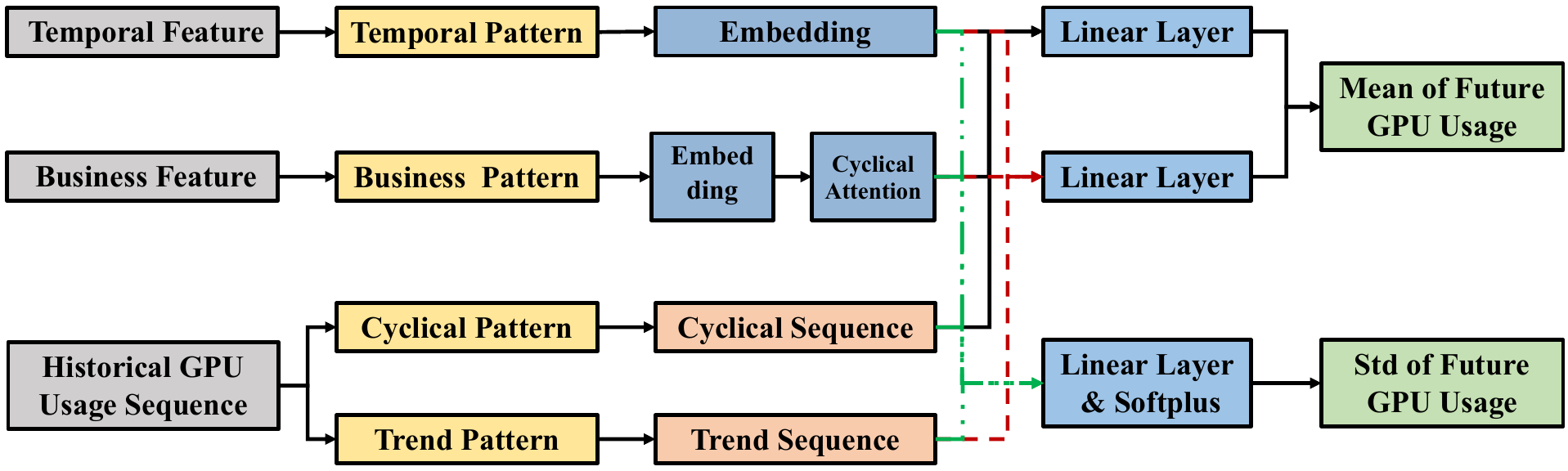}
    \vspace{-7mm}
    \caption{Framework of OrgLinear}
    \label{fig:orglinear_framework}
\end{figure}

GDE is the forecasting module of the GFS, providing reliable GPU demand projections for quota management and task scheduling. It generates a probability distribution for each organization’s GPU demand over future intervals by accounting for periodicity, bursts, and inter-organizational variations, capturing mean values and tail uncertainties. This demand distribution serves as the dynamic quota basis for the SQA and is also sent to the PTS to evaluate potential preemption costs.
However, accurate GPU demand prediction faces two challenges: (1) Temporal heterogeneity, where demand patterns exhibit organization-specific periodicity and volatility, and (2) Predictive uncertainty, as conventional point estimates do not adequately capture demand fluctuations essential for reliability-oriented scheduling. To address these issues, we propose OrgLinear, a hierarchical time-series forecasting model that integrates domain-specific decomposition with uncertainty quantification, as shown in Figure \ref{fig:orglinear_framework}.

\subsubsection{Adaptive Temporal Pattern Decomposition}

\label{para:periodicity}

The design of OrgLinear is informed by the temporal dynamics of GPU demand in production clusters. It exhibits two key properties: (1) Multi-scale periodicity - hierarchical cycles from hourly to weekly patterns coexist with varying dominance across organizations, and (2) Operational heterogeneity - demand fluctuations align with organization-specific business calendars that transcend common temporal cycles.  
As noted in Observation 2, the four organizations sharing A100 GPUs exhibit distinct multi-cycle behaviors. 
All display a diurnal periodicity (peaking from 10:00 to 24:00), but Organization C shows a significant weekly periodicity, with a 35.7\% drop in weekend demand, which is not observed in Organization A. These insights inspire our basic concept: Effective demand prediction requires the adaptive decomposition of shared periodic bases and organization-specific residual patterns.

Formally, let $\boldsymbol{\chi_o} \in \mathbb{R}^L$ denote the historical GPU usage sequence of size $L$ for organization $o$. Inspired by Autoformer \cite{Autoformer}, OrgLinear employs a \textit{domain-adaptive sliding kernel} to separate trend $\boldsymbol{\chi_{o, t}}$ and cyclical $\boldsymbol{\chi_{o, c}}$ components:
\begin{gather}
\boldsymbol{\chi_{o, t}} = \mathcal{K}_{\text{MA}}^d(\boldsymbol{\chi_o}) \\ \label{eq:time_series_resolve }\
\boldsymbol{\chi_{o, c}} = \boldsymbol{\chi_o} - \boldsymbol{\chi_{o, t}}
\end{gather}
where $\mathcal{K}_{\text{MA}}^d$ denotes a moving average kernel utilizing reflection padding to reduce boundary effects. In contrast to conventional methods using fixed basis functions, this data-driven decomposition adaptively captures organization-specific periodic signatures while maintaining temporal continuity, which is essential for addressing phase-shifted patterns. Table \ref{tab:notation} lists the commonly used notations in this work.

\begin{table}
\footnotesize
  \caption{Summary of commonly used notations}
  \label{tab:notation}
  \centering
  \begin{tabularx}{\linewidth}{>{\centering\arraybackslash}X l}
  \toprule
    Notation & Description\\ \midrule
     $\boldsymbol{\chi_o}$ & Historical GPU usage sequence of organization $o$ \\
     $\boldsymbol{\chi_{o,t}}$, $\boldsymbol{\chi_{o,c}}$ & Trend and cyclical components in $\boldsymbol{\chi_o}$\\
     $\boldsymbol{c_t}$ & Temporal feature at timestamp $t$\\
     $\boldsymbol{c_o}$ & Organization contextual feature\\
     $v_{oj}$ & $j$-dimensional business attributes of organization $o$ \\
     $H$ & GPU demand prediction horizon\\
     $\boldsymbol{\hat{y}_o}$ & Predicted GPU demand sequence of organization $o$\\
     $p$  & Target guarantee rate \\
     $\eta$ & Safety coefficient\\
     $e$  & Real spot eviction rate \\
     $C$ & Cluster GPU capacity\\
     $\mathbf{\mathcal{T}}$, $\mathbf{\mathcal{G}}$ & Task and node set\\     
    \bottomrule
  \end{tabularx}
\end{table}

Additionally, demand patterns vary with the time of day, day of the week, and holidays, making these factors crucial for accurate prediction. OrgLinear extracts the hour $t_{hour}$, weekday $t_{weekday}$, and holiday $t_{holiday}$ from the timestamp $t$. Each component is encoded into a feature vector of the same dimension via embedding layers. These vectors are then concatenated to form the temporal feature $\boldsymbol{c_t}$:
\begin{equation}
\boldsymbol{c_t} = Emb(t_{hour}) \oplus Emb(t_{weekday}) \oplus Emb(t_{holiday})
\end{equation}
where $\oplus$ denotes feature concatenation.

\subsubsection{Business Contextual Feature Extraction}

\begin{figure}
    \centering
    \includegraphics[width=0.85\linewidth]{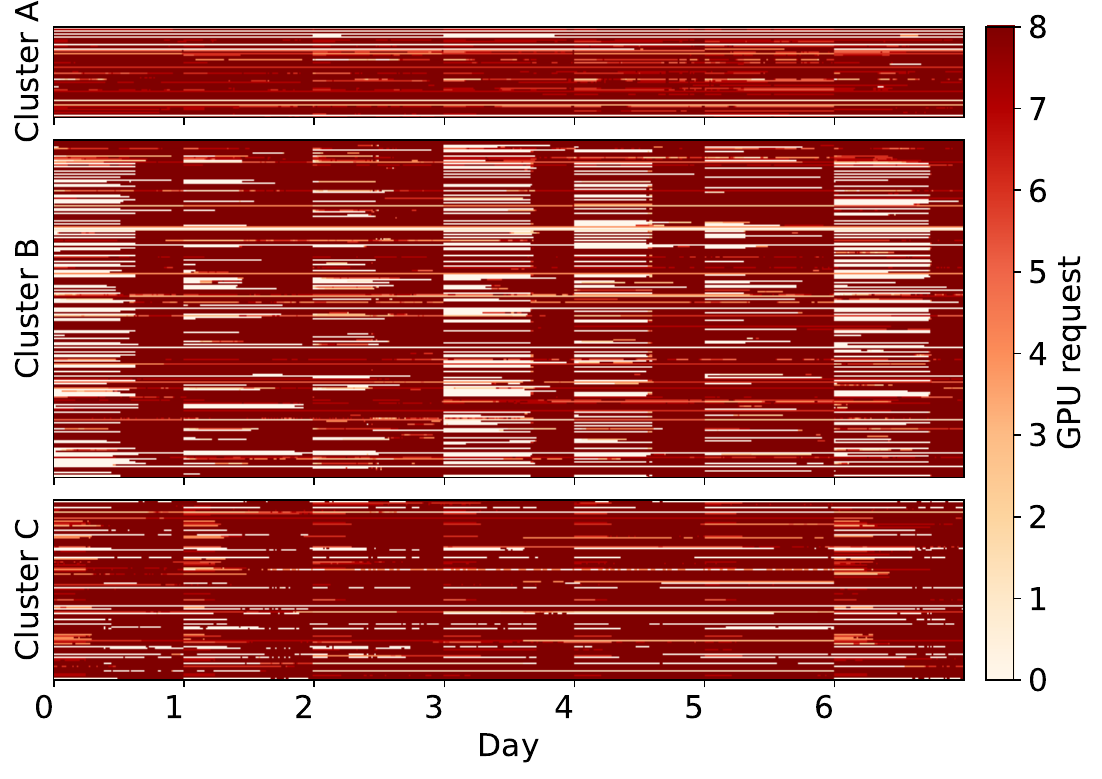}
    \vspace{-5mm}
    \caption{GPU allocation heatmap across 3 A100 clusters}
    \label{fig:part_util_heat}
    \vspace{-5mm}
\end{figure}

Beyond temporal dynamics, business metadata within traces provides crucial contextual information for GPU demand prediction, such as cluster affiliations and GPU model configurations. To quantify resource usage patterns, Figure \ref{fig:part_util_heat} presents a weekly heatmap visualization of node-level GPU allocation rates across three A100 clusters, where each contains about 500, 2,000 and 1,100 GPU cards, respectively. Each matrix represents a specific cluster, with rows indicating individual nodes, columns denoting 168 hourly time slots, and color intensity reflecting peak GPU allocation rates (0-8 cards per node-hour, based on each node's 8-card capacity).

The empirical analysis reveals notable inter-cluster heterogeneity in allocation patterns. Specifically, Cluster B exhibits a lower average GPU allocation rate (68.51\%) with more pronounced diurnal periodicity, particularly showing significant resource idleness during early morning hours. In contrast, Clusters A and C demonstrate comparatively higher allocation rates with less evident daily cyclical patterns. However, both Clusters A and C contain nodes that remain persistently idle throughout weekly operational cycles, indicating potential resource under-utilization at the node level.
Additionally, Spearman correlation \cite{de2016comparing} analysis indicates strong associations ($\rho> 0.68$, $p<0.01$) between cluster characteristics and the operational patterns of their affiliated organizations. This interdependence underscores the need for explicit modeling of cross-feature interactions among business attributes (e.g., cluster topological mappings, organization hardware preferences) to enhance demand prediction accuracy.

For each organization $o \in \mathcal{O}$, let $\boldsymbol{\mathcal{V}_o} = \{v_{o,1}, \cdots, v_{o,j}\}$ denote its $j$-dimensional business attributes (cluster, GPU model, etc.). We project each attribute $v_{o,k}$ into a dense representation space through learnable embedding matrices:
\begin{equation}
\boldsymbol{c_o} = Attention(Emb(v_{o,1}) \oplus Emb(v_{o,2}),\cdots,Emb(v_{o,j}))
\end{equation}

\subsubsection{Distribution-aware GPU Demand Forecasting}

OrgLinear formulates GPU demand forecasting as a probabilistic time series decomposition problem, where the objective is to learn latent temporal patterns and quantify prediction uncertainty through explicit normal distribution modeling. For each organization $o$, we model the future GPU demand $\boldsymbol{\hat{y}_o}$ over prediction horizon $H$ as $\boldsymbol{\hat{y}_o} \sim \mathcal{N}(\boldsymbol{\hat{\mu}_o}, \boldsymbol{\hat{\sigma}_o})$, where $\boldsymbol{\hat{\mu}_o}$ and $\boldsymbol{\hat{\sigma}_o}$ represent the predicted mean and standard deviation sequences, respectively.

Inspired by the linear decomposition paradigm \cite{sobieszczanski1982linear}, we implement two parallel linear projections for forecasting the cyclical $\boldsymbol{\hat{y}_{o,c}}$ and tread $\boldsymbol{\hat{y}_{o,t}}$ components in $\boldsymbol{\hat{y}_o}$:
\begin{equation}
\begin{aligned}
\boldsymbol{\hat{y}_{o,c}} &= \mathbf{W}_c^{\top}[\boldsymbol{\chi_{o,c}} \oplus \boldsymbol{c_o} \oplus \boldsymbol{c_t}] + \boldsymbol{b_c} \\\
\boldsymbol{\hat{y}_{o,t}} &= \mathbf{W}_t^{\top}[\boldsymbol{\chi_{o,t}} \oplus \boldsymbol{c_o} \oplus \boldsymbol{c_t}] + \boldsymbol{b_t} 
\end{aligned}
\end{equation}
where $\mathbf{W}_c$ and $\mathbf{W}_t$ are learnable weights. The final mean prediction is:
\begin{equation}
\boldsymbol{\hat{\mu}_o} = \boldsymbol{\hat{y}_{o,c}} + \boldsymbol{\hat{y}_{o,t}}
\end{equation}

\paragraph{Volatility Estimation with Variance Stabilization.} To capture time-varying uncertainty, we model demand volatility through heteroscedastic variance estimation \cite{chen2009conditional}:
\begin{equation}
\begin{aligned}
\boldsymbol{h_o} &= \mathbf{W}_v^{\top}[\boldsymbol{\chi_{o}} \oplus \boldsymbol{c_o} \oplus \boldsymbol{c_t}] + \boldsymbol{b_o} \\
\boldsymbol{\hat{\sigma}_o} &= \log(1 + \exp(\boldsymbol{h_o})) \quad 
\end{aligned}
\end{equation}
which ensures non-negative variance estimates while maintaining gradient stability. $\boldsymbol{h_o}$ describes the fluctuation of GPU demand of HP tasks.

\paragraph{Distributional Learning Objective}

The model parameters $\Theta = [\mathbf{W}_c,\mathbf{W}_t,\mathbf{W}_v,b_c,b_t,b_v]$ are optimized via maximum likelihood estimation \cite{myung2003tutorial}:
\begin{equation}
\mathcal{L}(\Theta) = -\sum_{o=1}^N \sum_{t=1}^H \log\phi\left(\frac{\boldsymbol{y_o^{(t)}} - \boldsymbol{\hat{\mu}_o^{(t)}}}{\boldsymbol{\hat{\sigma}_o^{(t)}}}\right)
\end{equation}
where $N$ is the number of organizations and $H$ is the prediction steps. $\boldsymbol{y_o^{(t)}}$ denotes the observed GPU demand for organization $o$ at future timestep $t$.

\subsection{Spot Quota Allocator (SQA)}

\label{sec:sqa}

The SQA translates the distributional demand forecasts from GDE into specific, time-varying quotas.
Its design focuses on two main objectives: (1) maintaining the SLOs of HP tasks during peak demand, and (2) enhancing idle resource usage through dynamic adjustments according to cluster conditions.

\subsubsection{GPU Inventory Estimation}
Our proposed temporal-spatial aggregation method operates in two complementary phases to aggregate organization demand predictions at the cluster level. First, we perform a temporal aggregation. For each organization $o$, we use the inverse cumulative distribution function (ICDF) of the Gaussian distribution to obtain a high-guarantee upper bound sequence of GPU demand, denoted as $\boldsymbol{\hat{y}_{o|p}}[1:H] = ICDF(p,\boldsymbol{\hat{\mu}_o},\boldsymbol{\hat{\sigma}_o})$, where $p$ is the target guarantee rate (e.g., 0.95, indicating that the eviction rate does not exceed 0.05).
Second, we sum the organization-specific peak demands across all $N$ organizations to compute the aggregate cluster requirements. This spatial composition prevents inter-tenant demand interference while maintaining global visibility of resource needs. Therefore, the GPU inventory with guaranteed $H$ hours $f(p,H)$ is estimated as: 
\begin{equation}
   \label{eq.guarantee}
   f(p,H) = C - \max(C, \sum_{o=1}^N \max(\boldsymbol{\hat{y}_{o|p}}[1:H]))
\end{equation}
where $C$ is the cluster GPU capacity. When the aggregated demand exceeds $C$, we set $f(p,H)=0$. This conservative choice prioritizes HP tasks by ensuring that no spot quota is allocated when demand is expected to saturate the cluster.

\subsubsection{Dynamic Quota Adjustment}

Based on the estimated GPU inventory, GFS partitions GPUs between HP and spot tasks. To reconcile long-term demand forecasts with current cluster status, under the premise that the guarantee rate is $p$ and the guarantee duration is $H$ hours, the available quota $Q_H$ for spot tasks in the future is defined as:
\begin{equation}
Q_H = \min\left(f(p, H)\cdot\eta,\ S_0 + S_a\right)
\label{eq:quota}
\end{equation}
\noindent
where $S_0$ denotes the current number of idle GPUs, $S_a$ is the number of allocated spot GPUs with a guaranteed duration of at least $H$ hours, and $\eta$ is a safety coefficient, initially set to 1.0, serving as a buffer to mitigate forecasting errors.

The core of SQA is the dynamic adjustment of $\eta$. The rationale is to continuously balance the trade-off between under-utilization (long queuing time for spot tasks) and over-allocation (high spot task evictions). To this end, $\eta$ is updated based on the cluster's recent conditions, specifically the real eviction rate $e$ and the maximum queuing time $l$ of spot tasks over the past $H$ hours. The update rule is defined as:
\begin{equation}
\label{eq:theta}
\eta_{\text{new}} = \begin{cases}
\eta_{\text{old}} \cdot \frac{p}{e}, & e > 1.5p, \\
\eta_{\text{old}} \cdot \left(1.5 - \frac{e}{p}\right), & e < 0.5p \text{ and } l > \theta, \\
\eta_{\text{old}}, & \text{otherwise}.
\end{cases}
\end{equation}
where $\theta$ is a preset threshold (e.g., 1 hour).

These rules consider three cases:
1) \textit{High Eviction.} When $e$ exceeds $1.5p$, it indicates that the current spot allocation may be too aggressive. Consequently, $\eta$ is reduced to lower spot quota and guarantee HP tasks. 
2) \textit{Low Eviction.} If $e$ is very low (i.e., less than $0.5p$) but queuing times are high ($l > \theta$), the system is likely being overly conservative. In this case, $\eta$ is increased to expand spot quota and reduce queuing delays. 
3) If neither condition is met, $\eta$ remains unchanged.

\subsection{Preemptive Task Scheduler (PTS)}

\label{sec:pts}

PTS is a crucial module that affects the efficiency of HP and spot tasks, as well as the overall cluster. It operates as the run–time component that converts high-level quotas from SQA into specific placement decisions on individual GPUs.  
Its objectives are twofold: (i) to ensure HP tasks receive the necessary resources and SLO compliance, and (ii) to minimize the cost of preempting spot tasks. Since most tasks require the same GPUs, PTS effectively manages heterogeneous clusters by scheduling tasks according to GPU models.

\subsubsection{Scheduling Optimization Problem Definition}
Scheduling GPU-intensive tasks faces several key challenges. (1) Tasks require varying GPU computing units (i.e., pods or instances). (2) A conflict exists between different types of tasks: HP tasks demand strict SLOs, whereas throughput-oriented spot tasks emphasize cost-efficient resource utilization. (3) To enhance fault tolerance, GPU tasks typically create multiple checkpoints during operation, allowing for recovery after a failure or preemption without starting over. Consequently, the scheduler must optimize GPU resource usage while minimizing the computational overhead associated with preemption or failures. To address these challenges, we establish a constrained optimization model that considers both eviction rate and cluster efficiency.

\label{para:problem_description}

Let $|\mathbf{\mathcal{T}}|$ tasks $\mathbf{\mathcal{T}} = \{\tau_1, \tau_2, \cdots, \tau_{|\mathbf{\mathcal{T}}|}\}$ submitted to $M$ nodes $\mathbf{\mathcal{G}} = \{n_1,n_2,\cdots,n_M\}$, where each node $n_j$ has $S_j$ GPUs. Each task $\tau_i=<w_i,g_i, \zeta_i,\mathbf{\psi}_i,\mathbf{\iota}_i>$ requests $w_i$ pods each containing $g_i$ GPUs, with task type $\zeta_i \in \{0, 1\}$ (spot/HP) and $D$ checkpoint milestones $\mathbf{\psi}_i = \{c_{i,1},\cdots,c_{i,D}\}$. Checkpoints are crucial for ensuring fault tolerance and efficient task resumption under potential preemption. 
Additionally, the set of $E_i$ runtime logs for $\tau_i$ is $\mathbf{\iota}_i = \{<t_{i,1}^s,t_{i,1}^e,f_{i,1}>,\cdots, <t_{i,E_i}^s,t_{i,E_i}^e,f_{i,E_i}>\}$. Each log signifies the $k$-th run of $\tau_i$, starting at time $t_{i,k}^s$ and ending at $t_{i,k}^e$, during which the task has achieved its $f_{i,k}$-th checkpoint. 
The scheduler must resolve the following dual objectives, which are formulated by mixed-integer linear programming (MILP) \cite{floudas2005mixed}:
\begin{subequations}
\footnotesize
\label{optimization_problem}
\begin{align}
\min 
\frac{\displaystyle \sum_{i\in|\mathcal{T}|}\sum_{j\in M} \sum _{t \in T}x_{i,j,t}(E_i-1)}{\displaystyle \sum_{i\in|\mathcal{T}|}\sum_{j\in M} \sum _{t \in T}x_{i,j,t}E_i}-\alpha\cdot\frac{\displaystyle \sum_{i\in|\mathcal{T}|}\sum_{j\in M} \sum _{t \in T}x_{i,j,t}w_ig_ic_{i,f_{i,E_i}}}{\sum_{j=1}^M S_jT}   \ \ \tag{\ref{optimization_problem}}
\end{align} 
\vspace{-5mm}
\begin{alignat}{2}
\text{s.t.}
& \sum_{i \in |{\mathcal{T}}|} x_{i,j,t}g_i \leq S_j,
&\label{resource capacity constraints}\\%
&  \sum_{j=1}^M x_{i,j,t} = w_i,\    &\label{Gang scheduling constraints}\\ 
&  (E_i-1)\zeta_i = 0 ,\   &\label{preemption constraints 1}\\ 
&  \sum_{j=1}^M x_{i,j,t_{i,k}^e} = 0,\   &\label{preemption constraints 2}\\ 
&  c_{i,f_{i,k}} - c_{i,f_{i,(k-1)}} \leq t_{i,k}^e - t_{i,k}^s < c_{i,(f_{i,k}+1)} - c_{i,f_{i,(k-1)}},\  \label{checkpoint constraints}  
\end{alignat}
\end{subequations}
where $x_{i,j,t}$ represents the number of pods assigned to task $\tau_i$ on node $n_j$ at time $t$ and $\alpha$ is the weighting factor. The constraint (\ref{resource capacity constraints}) enforces the physical GPU limits of each node. The constraint (\ref{Gang scheduling constraints}) indicates that the allocated resources need to meet the gang-scheduling. The constraints (\ref{preemption constraints 1}) and (\ref{preemption constraints 2}) restrict evictions to spot tasks only, mandating strict priority enforcement. Lastly, the constraint (\ref{checkpoint constraints}) ties task progress to completed checkpoints. 

The complexity of the problem manifests in several aspects. First, scheduling HP tasks presents significant challenges because they require varying resources while ensuring SLO guarantees, which is NP-hard \cite{sanders2022decentralized}. Additionally, scheduling spot tasks involves managing preemption strategies, further complicating the problem. Consequently, we propose a heuristic algorithm to address these challenges.

\subsubsection{Preemption-aware Scheduling Policy}

PTS adopts a preemptive scheduling strategy that efficiently manages cluster resources while processing new HP and spot tasks. 
The workflow of PTS is detailed in Figure \ref{fig:system_framework}. Prior to scheduling, PTS sorts tasks in the queue based on their GPU requests, pod requests, and submission times. During scheduling, PTS first attempts non-preemptive scheduling for HP tasks with available GPUs; if this fails, it switches to preemptive scheduling, allowing HP tasks to preempt spot tasks. Successful preemption results in the eviction of selected spot tasks.
For spot tasks, PTS strictly employs non-preemptive scheduling. Task deployment occurs only after successful resource allocation; otherwise, the task remains pending.

\begin{algorithm}[t]
\footnotesize
  \caption{Non-Preemptive Scheduling}
  \label{algo:job_placer}
        \KwIn{ Node set $\boldsymbol{\mathcal{G}}$, task $\tau_i=<w_i,g_i, \zeta_i,\psi_i>$}
        \KwOut{ Results $\boldsymbol{\mathcal{R}} = \{\cdots,x_{i,j,t}, \cdots\}$ or \textbf{Failed}}
        
        Initialize $\boldsymbol{\mathcal{R}}$ with $\emptyset$
        
        \While {$\sum_{j=1}^{\mid\boldsymbol{\mathcal{G}}\mid} x_{i,j,t} < w_i $}{
            
            Initialize candidate node set $\boldsymbol{\mathcal{G}_f}$ as $\emptyset$
            
            \If {$g_i < 1.0$} 
            { 
            $\boldsymbol{\mathcal{G}_f} \leftarrow \{ n_k \in \boldsymbol{\mathcal{G}} \mid \text{available GPUs of } n_k \geq g_i \}$
            }
            \Else
            {
            $\boldsymbol{\mathcal{G}_f} \leftarrow \{ n_k \in \boldsymbol{\mathcal{G}} \mid n_k.\text{idleGPUs} \geq g_i$ \text{ and } $Score_3(n_k) > 0 \}$
            }

            \If {$\mid \boldsymbol{\mathcal{G}_f} \mid > 0 $}{
                $n_k \leftarrow$ the top node in $\boldsymbol{\mathcal{G}_f}$ sorted by <Score$_1$, Score$_2$, Score$_3$> in descending
                
                $x_{i,k,t} \leftarrow x_{i,k,t} + 1$
                
                 }
            \Else{
                \Return \textbf{Failed}
            }
        }
        \Return $\mathcal{R}$
\end{algorithm}

\paragraph{Non-preemptive Scheduling}

\label{para:scheduler_score}

For a HP or spot task, PTS first filters nodes that meet the task's GPU requirements to form a candidate node set, then applies three evaluation criteria sequentially for each node to refine this set. These criteria effectively address GPU fragmentation, task SLO conflicts, and eviction risks. 


\textbf{Criterion 1: GPU Packing.} This involves prioritizing nodes with minimal idle GPU resources that meet the requirements. This approach helps prevent the inability to schedule subsequent tasks due to excessive fragmentation. Consequently, the scoring model is defined as:
\begin{equation}
Score_1 = Norm\left(1 - \frac{idleGPUs}{totalGPUs}\right) 
\label{equ:scheduler_score1}
\end{equation}
where $idleGPUs$ and $totalGPUs$ denote the number of idle GPUs and total GPUs on the node respectively, while $Norm()$ indicates the normalization operation.

\textbf{Criterion 2: Homogeneous Co-location.} Under Criterion 1, HP tasks are assigned to nodes running HP tasks, while spot tasks are directed to nodes that host spot tasks. This strategy primarily focuses on prioritizing the SLO of HP tasks and reduces the risk of HP tasks being unscheduled due to excessive fragmentation. By co-locating HP tasks, future HP tasks can be scheduled more efficiently by preempting spot tasks when necessary. Therefore, the scoring model considering workload-type co-location is expressed as:
\begin{equation}
Score_2 = \begin{cases}
Norm\left(\frac{hpGPUs}{totalGPUs}\right), & \text{for HP tasks} \\
Norm\left(\frac{spotGPUs}{totalGPUs}\right) , & \text{for Spot tasks}
\end{cases}
\label{equ:scheduler_score2}
\end{equation}
where $hpGPUs$ and $spotGPUs$ represent GPU allocations per workload type on the node.

\textbf{Criterion 3: Eviction Awareness.} When Criterion (1) and (2) fail to differentiate candidate nodes, the spot task is assigned to the node with the fewest evictions, while the HP task is allocated to the node with the highest past eviction count. This approach leverages historical resource patterns to decrease eviction chances, ensuring that HP and spot tasks are allocated according to past eviction trends, thus minimizing the risk of repeated evictions. The weighted node eviction rate $\bar{e}$ is defined as:
\begin{equation}
\bar{e} = \gamma \cdot e_{short} + (1-\gamma) \cdot \frac{e_{long}}{T_{long}}
\label{equ:eviction_frequency}
\end{equation}
where $\gamma$ balances short-term versus long-term eviction patterns. $T_{long}$ is the duration of long periods.
$e_{short}$ and $e_{long}$ denote the total number of eviction events in short and long periods. In the implementation, we define these periods as the past 1 hour and 24 hours, which are adjustable. The subsequent scoring applies asymmetric penalties:
\begin{equation}
Score_3 = \begin{cases}
\min(Norm(0.01m^{\bar{e}}), 1), & \text{HP tasks} \\
\max(Norm(1-0.01m^{\bar{e}}), 0), & \text{Spot tasks}
\end{cases}
\label{equ:scheduler_score3}
\end{equation}
where $m$ controls the penalty intensity, with increased values $\bar{e}$ that eliminate problematic nodes. When a node experiences high evictions ($Score_3=0$), a circuit breaker mechanism may activate. This mechanism automatically blacklists the node during the filter phase for spot task scheduling, preventing the repeated preemption of spot tasks on that node.

These scoring criteria provide a thorough evaluation of candidate nodes, enabling the scheduler to make informed resource allocation decisions. The non-preemptive scheduling procedure is outlined in Algorithm \ref{algo:job_placer}.

\paragraph{Preemptive Scheduling}

When a HP task cannot acquire resources through non-preemptive scheduling, it indicates insufficient idle GPUs. In such cases, the PTS enables the HP task to preempt spot tasks. However, this preemption may adversely impact the SLO of spot tasks. Thus, it is crucial to strategically select which spot tasks to preempt in order to minimize overall losses.

Preemptive scheduling differs from non-preemptive scheduling in two main aspects. First, it identifies candidate nodes by virtually preempting all currently running spot tasks, forming a candidate node set $\mathcal{G}_f$. Second, the score model employs a strategy focused solely on assessing preemption costs. As a result, the scheduler favors nodes with the lowest preemption costs for deploying HP tasks. This paradigm poses two challenges. First, it involves determining the optimal preemption combination, specifically, which GPU cards and spot tasks to preempt. Second, it requires a formal definition of preemption costs.

\begin{algorithm}[t]
\footnotesize
  \caption{Preemptive Scheduling}
  \label{algo:node_preempt}
    \KwIn {Node set $\boldsymbol{\mathcal{G}}$, HP task $\tau_i=<w_i,g_i, \zeta_i,\psi_i>$}
    \KwOut {Results $\boldsymbol{\mathcal{R}} = \{\cdots, x_{i,j,t}, \cdots\}$ or \textbf{Failed}}
    
    Initialize $\boldsymbol{\mathcal{R}}$ with  $\emptyset$
    
    \While {$\sum_{j=1}^{\mid\boldsymbol{\mathcal{G}}\mid} x_{i,j,t} < w_i $}{
     Initialize candidate node set $\boldsymbol{\mathcal{G}_f}$ as $\emptyset$
     
    \For{node $n_j$ in $\boldsymbol{\mathcal{G}}$}{
     Initialize GPU set $R$ and preempted spot task set $\mathcal{T}_j$ as $\emptyset$
     
    \For{each spot task $\tau_s$ running on node $n_j$}{
        $\mathcal{T}_j\leftarrow \mathcal{T}_j \cup \{\tau_s\}$, $R \leftarrow R \cup \tau_s.GPUs$
    }
    
    Sort $\mathcal{T}_j$ by descending waste time $\vartheta_{\tau_s}$ (Eq.~\ref{equ:waste_time})
    
    \For {$\tau_s \in \mathcal{T}_j$}{
        \If {$(R - \tau_s.GPUs)$ can satisfy $\tau_i$'s GPU requirements}{
            $R \leftarrow R - \tau_s.GPUs$, $\mathcal{T}_j \leftarrow \mathcal{T}_j 
 - \tau_s$
        }
    }
    $\textbf{X}_j \leftarrow \text{Minimal GPU set satisfying } \tau_i \text{ in } R$
    
    
     Add tuple $<n_j, \mathcal{T}_j>$ to $\boldsymbol{\mathcal{G}_f}$
    }
     \If {$\mid \boldsymbol{\mathcal{G}_f} \mid > 0 $}{
         Select node $n_k \in \boldsymbol{\mathcal{G}_f}$ with minimum $cost(n_k)$
        
         $x_{i,k,t} \leftarrow x_{i,k,t} + 1$
         }
    \Else{
        \Return \textbf{Failed}
    }
    }
    \Return $\mathcal{R}$
\end{algorithm}

The preemption scheduling (Algorithm \ref{algo:node_preempt}) employs a waste-aware heuristic. It obtains the set of all spot tasks $\mathcal{T}_k$ running on all candidate nodes in $\mathcal{G}_f$ (lines 5-7). For each spot task $\tau_s \in \mathcal{T}_k$, we calculate the resource waste due to preemption, resulting in unsaved states:
\begin{equation}
\vartheta_{\tau_s} = g_{\tau_j} \cdot (t - t^{check}_{\tau_s})
\label{equ:waste_time}
\end{equation}
where $g_{\tau_s}$ denotes the requested GPUs of $\tau_s$, and $(t - t^{check}_{\tau_s})$ represents the elapsed time since last checkpoint. By sorting candidates in descending order of wastes (line 8), the algorithm prioritizes preserving high-value tasks that are likely nearing the completion of guaranteed hours (checkpoints). Moreover, the linear scan (lines 9–13) ensures $O(|\mathcal{T}_k|)$ complexity while achieving near-optimal solutions.
The final GPU selection (lines 14-18) aims to ensure that the spot tasks in $\mathcal{T}_k$ running on node $n_k$ with the lowest preemption cost are chosen. The preemption cost for node $n_k$ is calculate by:
\begin{equation}
cost(n_k) = \underbrace{\frac{F+|\mathcal{T}_k|}{G+F+|\mathcal{T}_k|} - \frac{F}{G+F}}_{\text{Eviction rate impact}} + \beta\cdot\underbrace{\frac{\sum_{\tau_s\in\mathcal{T}_k} \vartheta_{\tau_s}}{\sum_{k=1} ^M S_k\cdot T}}_{\text{Usage impact}}
\label{equ:delta_derivation}
\end{equation}
where $G$ and $F$ represent the number of historical successful and evicted spot tasks, respectively in the cluster, and $S_k\cdot T$ denotes the total execution time of GPUs in node $n_k$. Since the term $-\frac{F}{G+F}$ in Eq. (\ref{equ:delta_derivation}) is a constant for total nodes, the simplified cost metric becomes:
\begin{equation}
cost(n_k) = \frac{F+|\mathcal{T}_k|}{G+F+|\mathcal{T}_k|} + \ \beta\cdot\frac{\sum_{\tau_s \in \mathcal{T}_k} \vartheta_{\tau_s}}{\sum_{k = 1} ^M S_k\cdot T}
\label{equ:preempt_cost}
\end{equation}

PTS selects nodes with the minimal $cost(n_k)$, balancing eviction rates and resource usage through $\beta$. This design ensures that the system remains aligned with the overarching optimization objective. 
Finally, the entire process of PTS is outlined in Algorithm \ref{algo:PTS}. The time complexity of Algorithm \ref{algo:job_placer} is $O(\mid w_i\mid  \mid \mathcal{G} \mid)$, and Algorithm \ref{algo:node_preempt} is $O(\mid w_i\mid \mid \mathcal{G} \mid  \mid \mathcal{T}_k \mid)$, so PTS's time complexity is $O(\mid w_i\mid  \mid \mathcal{G} \mid  \mid \mathcal{T}_k \mid)$. The average time to schedule a task is < 1 second in our implementation.

\begin{algorithm}[t]
\footnotesize
  \caption{Preemptive Task Scheduler (PTS)}
  \label{algo:PTS}
    \KwIn {Node Set $\mathcal{G}$, task $\tau_i=<w_i,g_i,\zeta_i>$, time $t$}
    \KwOut{ Results $\mathcal{R} = \{\cdots, x_{i,j,t}, \cdots\}$ or \textbf{Failed}}
    \If{\textbf{not} $Quota\_satisfy(\tau_i)$}{
        \Return \textbf{Failed}
    }
     $\mathcal{R} \leftarrow $ Non-Preemptive Scheduling $(\mathcal{G}, \tau_i, t)$ (Alg. \ref{algo:job_placer})
     
    \If{$\mathcal{R} = \textbf{Failed}$ \textbf{and} $\zeta_i = 1$ }{
        $\mathcal{R} \leftarrow $ Preemptive Scheduling $(\mathcal{G}, \tau_i, t)$ (Alg. \ref{algo:node_preempt})
    }
    \Return $\mathcal{R}$
\end{algorithm}

\section{Performance Evaluation}
\label{sec:experiment}



\subsection{Setup}

\label{para:experiment_setup}

We deployed GFS in large-scale GPU production clusters and verified its effectiveness through simulations with real-world traces. 
The traces involved 1 cluster with 2,296 A100 GPUs (287 8-card nodes) and utilized various real workload patterns, including partial, single-card, multi-card, full-node 8-card, and gang requests.

\label{para:dataset}

\begin{table}[htp]
\footnotesize
  \caption{An overview of two types of tasks}
  \label{tab:job_dataset}
  \vspace{-4mm}
  \centering
  \begin{tabular}{@{}ccccccccc@{}} \toprule
    & & \multicolumn{5}{c}{Request GPU Specification(\%)} & \multicolumn{2}{c}{Gang(\%)} \\ \cmidrule(r){3-7} \cmidrule(r){8-9}
    Type &Number & <1 & 1 &2&4 & 8 & Yes & No\\ \midrule
    HP &  138,403 & 0.11 & 55.11 & 13.37 & 7.53 & 23.69 & 8.66 & 91.34 \\
    Spot & 26,635 & 0.82 & 67.35 & 5.67 & 12.00 & 14.04 & 27.26 & 72.74 \\ \bottomrule
  \end{tabular}
  \vspace{-2mm}
\end{table}

\textbf{Dataset.} The dataset includes all HP and spot tasks submitted from Apr to Jun 2024. As shown in Table \ref{tab:job_dataset}, HP tasks represent the majority (83.86\% of total tasks), whereas spot tasks make up 16.14\%. Most tasks typically request 1, 2, 4, or 8 full cards, with 1-card and 8-card requests being the most prevalent. Importantly, 8.66\% of HP tasks and 27.26\% of spot tasks require gang scheduling. 

\textbf{Workloads.} We assess three distinct spot workloads, aligning HP tasks with actual datasets. (1) \textbf{Low Spot Workload}: Spot tasks retain their original submission rate. (2) \textbf{Medium Spot Workload}: Spot tasks are scaled to 200\% of their original submission rate. (3) \textbf{High Spot Workload}: Spot tasks are scaled to 400\%.

\textbf{Baselines.}
We compare GFS with four schedulers. (1) \textbf{YARN-CS} \cite{YARN} is a classic scheduling framework utilizing FCFS management and a best-fit strategy, supporting preemption. (2) \textbf{Chronus} \cite{Chronus} utilizes MILP and local-search to allocate time-limited leases to SLO and Best-effort tasks. Tasks are guaranteed within the lease period, although preemption is not supported. In our implementation, HP and spot tasks correspond to SLO and Best-effort tasks, with lease periods of 20 and 5 minutes, respectively. (3) \textbf{Lyra} \cite{Lyra} leases idle inference nodes to training tasks and employs a heuristic to minimize preemption costs. HP and spot tasks are mapped to inference and training tasks. The algorithm reduces preemption costs of training tasks caused by inference tasks, similar to our work. (4) \textbf{FGD} \cite{FGD_atc23} schedules requests to minimize fragmentation, optimizing GPU sharing efficiency. Our implementation adapts the method for quantifying GPU fragmentation from in-card to in-node, aiming to reduce fragmentation and accurately replicate the original design.

\begin{table}[tb]
\footnotesize
  \caption{Parameter settings of GFS}
  \vspace{-4mm}
  \label{tab:hyperparameter}
  \centering
  \begin{tabularx}{\linewidth}{lc}
  \toprule
    Symbol & Specification \\ \midrule
    Weight coefficient $\alpha$ in Eq. (\ref{optimization_problem}
    ) & 0.5 \\
    Weight coefficient $\beta$ in Eq. (\ref{equ:preempt_cost})  & 0.5 \\
    Target guarantee rate $p$ (\%) in Eq. (\ref{eq.guarantee})  & 0.9 \\
    Maximum JQT threshold $\theta$ (s) in Eq. (\ref{eq:theta})& 3,600 \\
    Weight ratio $\gamma$ in Eq. (\ref{equ:eviction_frequency})& 0.8 \\
    Penalty $m$ in Eq. (\ref{equ:scheduler_score3}) & 3 \\
    Guarantee hours $H$ in Eq. (\ref{eq.guarantee})& [\textbf{1}, 2, 4]  \\
    Spot quota update interval (s)  & 300  \\
    \bottomrule
  \end{tabularx}
  \vspace{-2mm}
\end{table}

\textbf{Parameters.} The parameter settings of GFS are listed in Table \ref{tab:hyperparameter}, which are selected based on existing industry practices \cite{EC2_Spot, GCE} and theoretical considerations \cite{han2024inss}.

\subsection{Evaluation Metrics}

\textbf{Job Completion Time (JCT).} The completion time of a task is the difference between its finish and submission times. JCT denotes the average completion time for a set of tasks and is a key metric for evaluating scheduler performance \cite{sun2021deepweave}.

\noindent
\textbf{Job Queuing Time (JQT).} The queuing time of a task is the interval between its entry into the waiting queue and the start of its execution. For spot tasks, the queuing time may comprise multiple segments if preempted. In these instances, the total queuing time is the cumulative sum of all segments.

\noindent
\textbf{Eviction Rate ($e$).} This metric is defined as the ratio of the number of times a spot task is evicted to the number of times it runs. The eviction rate indicates the SLO guarantee for spot tasks. The eviction rate of HP tasks remains 0.

\subsection{Production Cluster Deployment}

\begin{figure}[tb]
    \centering
    \subfloat[Spot Eviction(\%)]{
    \includegraphics[width = 0.45\linewidth]{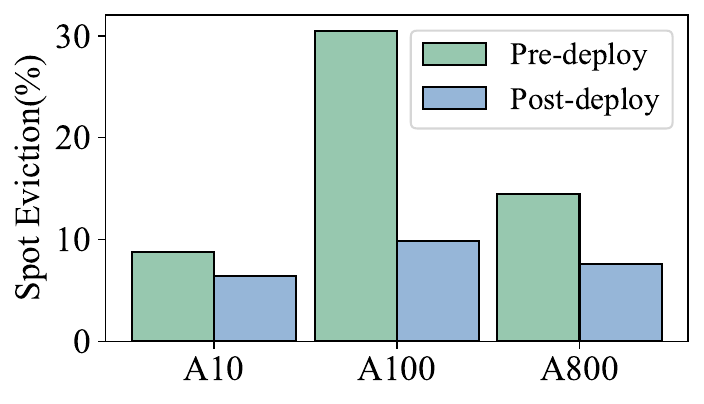}
    \label{fig:spot_eviction}
  }
  \hspace{2pt}
    \subfloat[GPU Allocation Rate(\%)]{
    \includegraphics[width=0.45 \linewidth]{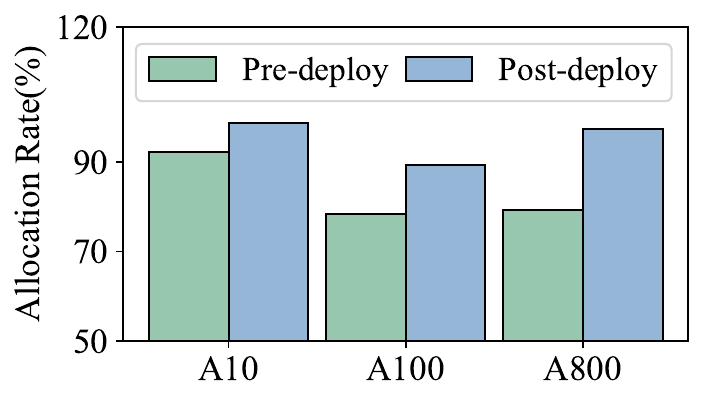}
     \label{fig:allocation_rate}
    }
    \vspace{-3mm}
    \caption{Deployment of the GFS in large-scale clusters}
    \label{tab:production_results}
    \vspace{-7mm}
\end{figure}

Figure \ref{tab:production_results} compares performance metrics before (Jan 2024) and after (Oct 2024) deploying GFS in a production cluster with the largest number of GPUs, as detailed in Table \ref{tab:node_preview}. In Oct 2024, tenants submitted a total of 541,207 HP tasks and 114,226 spot tasks. The primary focus is on the eviction rate of spot tasks and the GPU allocation rates.
Post-deployment results show significant improvements across all GPU models, with the eviction rate of spot tasks consistently falling below 10\%. Specifically, the A100 model achieved an impressive 67.81\% reduction in eviction rate.

In terms of GPU allocation rate, significant gains were noted in Figure \ref{fig:allocation_rate}. The A800 GPU model exhibited the highest improvement, with a 22.79\% increase, indicating more efficient resource utilization. Even the A10 model, which had a lower eviction rate prior to deployment, saw a notable rise of 6.94\% in GPU allocation rate, achieving an impressive 98.68\%. The A100 model's allocation rate increased by 14.03\%. Overall, GFS improves approximately \$459,715 in monthly gains, based on cloud GPU pricing \cite{Ali_price}.

\subsection{Scheduling Simulation Comparison}
The three performance metrics for HP/spot tasks are presented in Table \ref{tab:scheduler_results}, with each subtable representing a distinct workload scenario. 
The results show that, regardless of the spot workloads, GFS consistently outperforms the four baselines across all metrics for meeting the SLOs of both task types. For HP tasks, while ensuring JCT, GFS reduces their average JQT by $60.17\%-70.81\%$, thanks to preemption optimization. On average, HP tasks experience a wait time of only 28.8 seconds. Furthermore, the P99th JCT remains unchanged across all spot workloads, indicating that tail latency is immune to spot workload fluctuations. This robustness stems from the fact that GFS performs strict priority scheduling, ensuring that HP tasks are immediately and exclusively allocated resources, regardless of fluctuations in Spot workloads.

\begin{table}
\centering
\footnotesize
\setlength{\tabcolsep}{4pt}
  \caption{Scheduling metrics comparison with four baselines}
  \label{tab:scheduler_results}
  \centering
  \begin{subtable}{\linewidth}
  \centering
  \caption{Low Spot Workload}
  \label{tab:low_scheduler_results}
      \begin{tabular}{@{}ccccccc@{}} \toprule
        &  \multicolumn{3}{c}{HP tasks} & \multicolumn{3}{c}{Spot tasks} \\ \cmidrule(r){2-4} \cmidrule(r){5-7}
        &JCT-p99(s) &JCT(s) & JQT(s) & JCT(s) & JQT(s) & $e$(\%)  \\ \midrule
        YARN-CS &  29,308.5 & 17,865.9 & 77.4 & 13,287.4 & 3,110.2 & 2.32   \\
        Chronus & 32,580.0 & 19,925.8 & \underline{71.4} & \underline{12,068.2} & \underline{724.7} & - \footnotemark \\
        Lyra &\underline{29,304.5}  & \underline{17,863.5} & 114.9 & 12,473.2 & 2,360.9 & \underline{1.30}   \\
        FGD & 29,308.5 &17,905.3 & 116.8 & 15,680.9 & 4,887.7 & 17.32   \\
        GFS &\textbf{29,304.5}  &\textbf{17,777.1} & \textbf{28.4} & \textbf{10,438.7} & \textbf{323.0} & \textbf{0.74} \\ \midrule
        IMP. & 0.00\% & 0.48\% & 60.17\% & 13.50\% & 55.43\% & 43.08\%  \\
        \bottomrule
    \end{tabular}
  \end{subtable}

  \begin{subtable}{\linewidth}
      \centering
      \caption{Medium Spot Workload}
      \label{tab:medium_scheduler_results}
      \begin{tabular}{@{}ccccccc@{}} \toprule
        &  \multicolumn{3}{c}{HP tasks} & \multicolumn{3}{c}{Spot tasks} \\ \cmidrule(r){2-4} \cmidrule(r){5-7}
        &JCT-p99(s) &JCT(s) & JQT(s) & JCT(s) & JQT(s) & $e$(\%)  \\ \midrule
        YARN-CS & 29,308.5 &  17,891.5 & 102.9 & 15,866.3 & 5,087.4 & 16.74   \\
        Chronus & 32,580.0 & 19,930.0 & \underline{76.0} & \underline{12,538.4} & \underline{1,211.7} & -  \\
        Lyra & \underline{29,304.5} & \underline{17,862.3} & 113.7 & 15,107.9 & 5,016.8 & \underline{1.78}   \\
        FGD & 29,308.5 & 17,900.0 & 111.5 & 16,168.0 & 5,450.5 & 15.36   \\
        GFS & \textbf{29,304.5} & \textbf{17,776.7} & \textbf{27.7} & \textbf{10,715.3} & \textbf{575.4} & \textbf{1.21} \\ \midrule
        IMP. & 0.00\% & 0.48\% & 63.53\% & 14.54\% & 48.70\% & 32.02\%  \\
        \bottomrule
    \end{tabular}
  \end{subtable}

  \begin{subtable}{\linewidth}
      \centering
      \caption{High Spot Workload}
      \label{tab:high_scheduler_results}
     \begin{tabular}{@{}ccccccc@{}} \toprule
        &  \multicolumn{3}{c}{HP tasks} & \multicolumn{3}{c}{Spot tasks} \\ \cmidrule(r){2-4} \cmidrule(r){5-7}
        &JCT-p99(s) &JCT(s) & JQT(s) & JCT(s) & JQT(s) & $e$(\%)  \\ \midrule
        YARN-CS & 29,308.5 &  17,892.3 & \underline{103.9} & 17,221.1 & 6,494.5 & 14.94   \\
        Chronus & 32,580.0 & 20,084.3 & 230.7 & \underline{15,373.0} & \underline{4,039.2} & -  \\
        Lyra & \underline{29,304.5} & \underline{17,867.8} & 119.1 & 21,123.7 & 11,026.5 & \underline{1.63}   \\
        FGD & 29,308.5 & 17,901.0 & 112.5 & 20,741.3 & 10,110.6 & 12.99   \\
        GFS & \textbf{29,304.5} & \textbf{17,780.3} & \textbf{30.3} & \textbf{13,117.1} & \textbf{2,901.0} & \textbf{1.24} \\ \midrule
        IMP. & 0.00\% & 0.49\% & 70.81\% & 14.67\% & 28.18\% & 23.93\%  \\
        \bottomrule
    \end{tabular}
  \end{subtable}
  \vspace{-3mm}
\end{table}
\footnotetext{Chronus employs lease-based scheduling, allowing preemption solely at lease expiration. Thus, its eviction rate is omitted from statistics.}

For spot tasks, GFS achieves significant improvements, reducing JCT by $14.24\%$, shortening JQT by $44.10\%$, and lowering eviction rate by 33.01\% on average. Notably, Chronus, the second-best scheduler, compromises HP task performance, with its JCT for HP tasks being 11.34\%-12.11\% higher than that of other schedulers, to achieve a shorter JCT for spot tasks. Lyra maintains a low eviction rate (1.78\%) but faces challenges with long JQT due to inefficient scoring. YARN-CS and FGD perform poorly across all metrics.

\subsection{Parameters Sensitivity Analysis}

We performed a sensitivity analysis on the guarantee hours ($H$), set at 1, 2, and 4 hours, to evaluate its effect on spot task assurance performance under medium workloads. As indicated in Table \ref{tab:sensitivity_analysis}, when $H$ is 1 or 2 hours, the SLO of spot tasks remains largely stable. However, extending $H$ to 4 hours reduces the available quota of spot tasks that can be reliably guaranteed against preemption throughout the entire period. This decrease leads to increased queuing times for spot tasks, resulting in longer JCT. Importantly, the spot task eviction rate stays low (<1.5\%) across all configurations.


\begin{table}[tb]
\footnotesize
  \caption{The sensitivity analysis of guarantee hours $H$}
  \label{tab:sensitivity_analysis}
  \vspace{-3mm}
  \centering
  \begin{tabular}{@{}cccccc@{}} \toprule
    &  \multicolumn{2}{c}{HP tasks} & \multicolumn{3}{c}{Spot tasks} \\ \cmidrule(r){2-3} \cmidrule(r){4-6}
     $H$ &JCT(s) & JQT(s) & JCT(s) & JQT(s) & $e$(\%)  \\ \midrule
    1 & 17,776.7 & 27.7 & 10,715.3 & 575.4 & 1.21 \\ 
    2 & 17,776.9 & 28.4 & 10,619.0 & 506.6 & 1.29 \\ 
    4 & 17,776.9 & 28.4 & 13,252.8 & 3,145.6 & 1.16 \\ 
    \bottomrule
  \end{tabular}
  \vspace{-4mm}
\end{table}

\subsection{Ablation Studies}

\subsubsection{GPU Demand Estimator}
We compare OrgLinear with six state-of-the-art time-series forecasting baselines. (1) Four \textbf{Transformer-based} models, including Transformer\cite{vaswani2023attentionneed}, Informer \cite{zhou2021informerefficienttransformerlong}, Autoformer \cite{Autoformer}, and FEDformer \cite{zhou2022fedformerfrequencyenhanceddecomposed}, are effective in capturing long-range dependencies. (2) \textbf{DLinear} \cite{zeng2022transformerseffectivetimeseries} is a linear baseline that decomposes sequences into trend and periodic components. (3) \textbf{DeepAR} \cite{salinas2019deeparprobabilisticforecastingautoregressive} is an LSTM-based probabilistic model capable of quantile prediction.

We use four metrics: Mean Absolute Error (MAE), Mean Squared Error (MSE), Root Mean Squared Error (RMSE), and Mean Absolute Percentage Error (MAPE) to measure the accuracy. Additionally, we introduce Mean Absolute Quantile Error (p-MAQE), which measures the average absolute error between predicted and actual values at a specific quantile level (p), to assess the effectiveness of OrgLinear.

\begin{figure*}[tb]
    \centering
    \includegraphics[scale=0.5]{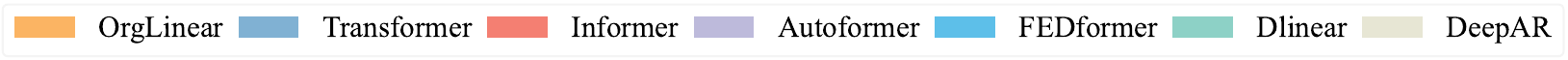}

    \vspace{-1mm}
  \subfloat[MAE]{
    \includegraphics[scale=0.4]{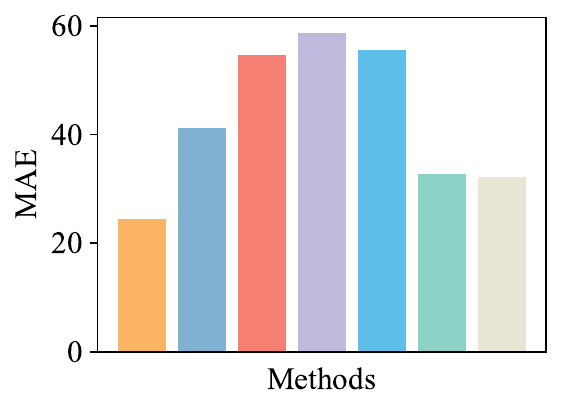}
        \label{fig:mae}
  }
  \subfloat[MSE]{
    \includegraphics[scale=0.4]{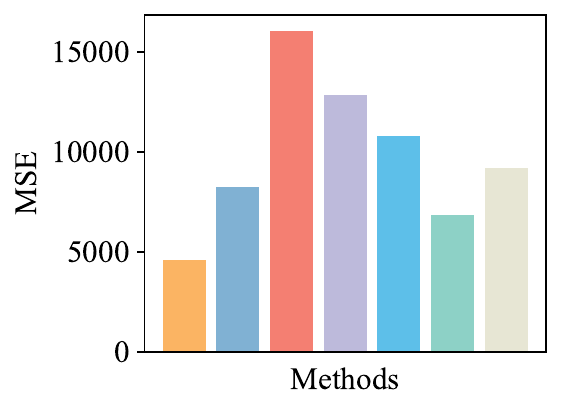}
        \label{fig:mse}
  }
  \subfloat[RMSE]{
    \includegraphics[scale=0.4]{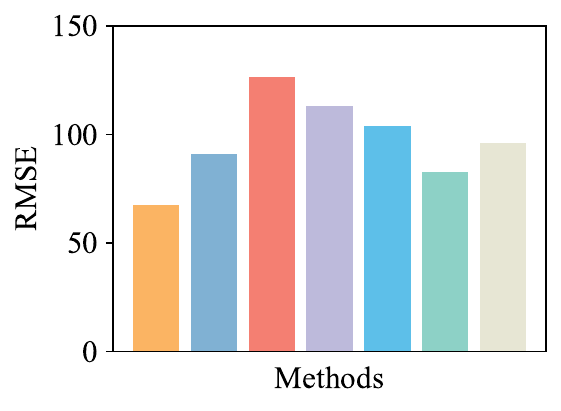}
        \label{fig:rmse}
  }
  \subfloat[MAPE]{
    \includegraphics[scale=0.4]{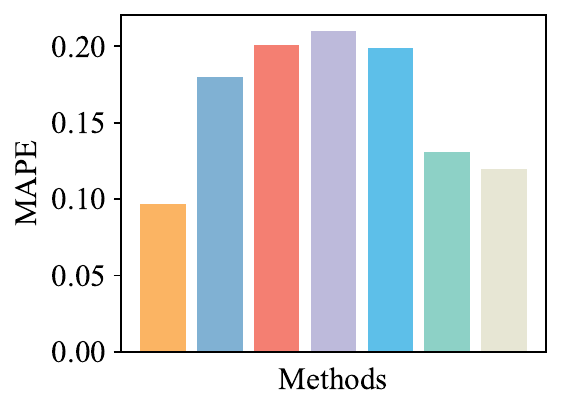}
        \label{fig:mape}
    
    }
  \vspace{-0.3cm}
\caption{Model performance comparison among OrgLinear and baselines}
\label{fig:orglinear}
\vspace{-4mm}
\end{figure*}


Figure \ref{fig:orglinear} demonstrates that OrgLinear significantly outperforms the six baselines. Specifically, OrgLinear achieves a reduction of 24.30\% in MAE, 33.26\% in MSE, 18.31\% in RMSE, and 19.17\% in MAPE compared to the suboptimal. This improvement can be attributed to OrgLinear's effective utilization of business and temporal features, along with its application of linear models for predicting probability distributions.
Furthermore, we assessed OrgLinear’s quantile prediction accuracy at $p=0.9$ and $p=0.95$ compared to DeepAR (the top baseline), as illustrated in Table \ref{tab:prob_evaluation}. The findings indicate that OrgLinear significantly surpasses DeepAR, achieving improvements of 45.61\% in 0.95-MAQE and 28.21\% in 0.9-MAQE. Additionally, OrgLinear's training time is only 1.63\% of DeepAR's, making it highly suitable for real-time GPU demand forecasting in large-scale clusters.

\begin{table}
\footnotesize
  \caption{Evaluation of quantile prediction accuracy}
  \label{tab:prob_evaluation}
  \vspace{-3mm}
  \centering
  \begin{tabular}{@{}c|cc|c@{}} 
  \toprule
       & 0.95-MAQE & 0.9-MAQE & Training Time(s) \\ \midrule
      DeepAR & 0.057 & 0.078 & 28,610.1  \\
      OrgLinear & 0.026 & 0.056 & 464.9 \\
    \bottomrule
  \end{tabular}
  \vspace{-3mm}
\end{table}

\begin{table}[tp]
\footnotesize
  \caption{The ablation experimental results of GDE}
  \label{tab:estimator_ablation_results}
  \vspace{-3mm}
  \centering
  \begin{tabular}{@{}cccccc@{}} \toprule
    &  \multicolumn{2}{c}{HP tasks} & \multicolumn{3}{c}{Spot tasks} \\ \cmidrule(r){2-3} \cmidrule(r){4-6}
    &JCT(s) & JQT(s) & JCT(s) & JQT(s) & $e$(\%)  \\ \midrule
    GFS-e & 17,776.2 & 27.4 & 20,687.1 & 10,502.4 & 8.08  \\
    GFS & 17,776.7 & 27.7 & 10,715.3 & 575.4 & 1.21 \\ 
    \bottomrule
  \end{tabular}
  \vspace{-3mm}
\end{table}

Furthermore, we assess the effectiveness of GDE. The baseline comparison employs GFS-e, which preserves the same SQA and PTS as GFS but replaces the GPU demand prediction algorithm in GDE with a method that takes the peak GPU demand of tenants from the previous week for future predictions. This was a naive and conservative approach adopted in the production cluster. Table \ref{tab:estimator_ablation_results} shows the ablation results of GDE. Although GDE has a minimal effect on SLO compliance for HP tasks, it significantly improves SLO for spot tasks, decreasing JCT, JQT, and $e$ by 48.20\%, 94.52\%, and 85.02\%, respectively, compared to the baseline. These metrics underscore GDE's critical role in enhancing SLO guarantees for spot tasks and scheduling efficiency in GFS.


\subsubsection{Spot Quota Allocator}

\begin{table}[tp]
\footnotesize
  \caption{The ablation experimental results of SQA}
  \label{tab:divider_ablation_results}
  \vspace{-3mm}
  \centering
  \begin{tabular}{@{}cccccc@{}} \toprule
    &  \multicolumn{2}{c}{HP tasks} & \multicolumn{3}{c}{Spot tasks} \\ \cmidrule(r){2-3} \cmidrule(r){4-6}
    &JCT(s) & JQT(s) & JCT(s) & JQT(s) & $e$(\%)  \\ \midrule
    GFS-d & 17,776.6 & 27.6 & 12,304.7 & 2,174.3 & 1.73  \\
    GFS & 17,776.7 & 27.7 & 10,715.3 & 575.4 & 1.21 \\ 
    \bottomrule
  \end{tabular}
  \vspace{-5mm}
\end{table}
We assess the effectiveness of SQA. The baseline comparison employs GFS-d, which maintains the same GDE and PTS as GFS but sets $\eta = 1.0$ in Eq. (\ref{eq:quota}) to separate quota allocation from real-time cluster feedback adaptation. 
Table \ref{tab:divider_ablation_results} shows the ablation results of SQA. Although SQA has a minimal effect on SLO compliance for HP tasks, it significantly improves SLO for spot tasks, decreasing JCT, JQT, and $e$ by 12.92\%, 73.53\%, and 30.06\%, respectively, compared to the baseline. These metrics highlight SQA's essential contribution to enhancing SLO guarantees for spot tasks and scheduling efficiency in GFS.

\subsubsection{Preemptive Task Scheduler}

To validate the efficacy of the two core modules in PTS, we compare GFS with three degraded variants. (1) \textbf{GFS-s}: Replaces the non-preemptive scheduling module with a simplified version that only considers GPU stacking (equivalent to best fit), disabling homogeneous co-location and eviction awareness. (2) \textbf{GFS-p}: Replaces the preemptive scheduling module with random node/card selection. (3) \textbf{GFS-sp}: Combines both degraded modules from GFS-s and GFS-p.

Table \ref{tab:scheduler_ablation_results} shows that both modules significantly enhance spot task performance. Upgrading GFS-sp to GFS-p through the reintroduction of preemptive scheduling decreases JCT for spot tasks by 10.99\%. In parallel, transitioning from GFS-sp to GFS-s by reinstating non-preemptive scheduling results in an 11.71\% reduction in JCT. The combination of both optimizations (GFS) yields a total reduction of 23.52\%. 
Additionally, restoring the non-preemptive module leads to a 40.71\% decrease in JQT. Likewise, reinstating the preemptive module achieves a 41.29\% reduction by minimizing re-queuing delays associated with preemption.

\begin{table}[tbp]
\footnotesize
  \caption{The ablation results of PTS}
  \label{tab:scheduler_ablation_results}
  \vspace{-4mm}
  \centering
  \begin{tabular}{@{}cccccc@{}} \toprule
    &  \multicolumn{2}{c}{HP tasks} & \multicolumn{3}{c}{Spot tasks} \\ \cmidrule(r){2-3} \cmidrule(r){4-6}
    &JCT(s) & JQT(s) & JCT(s) & JQT(s) & $e$(\%)  \\ \midrule
    GFS-sp & 17,778.1 & 29.3 & 14,010.5 & 3,811.5 & 2.64  \\
    GFS-s & 17,776.8  & 27.7 & 12,369.7 & 2,237.7 & 1.79  \\
    GFS-p & 17,776.2 & 27.3 & 12,471.0 & 2,259.7 & 2.95   \\
    GFS & 17,776.7 & 27.7 & 10,715.3 & 575.4 & 1.21 \\ 
    \bottomrule
  \end{tabular}
  \vspace{-5mm}
\end{table}

\section{Related Work}

\label{sec:relatedwork}

\subsection{Spot Instances in Cloud Computing}

Spot instances are widely used by cloud providers like Amazon \cite{EC2_Spot}, Google \cite{GCE}, and Microsoft \cite{Azure_Spot} to reduce cloud computing costs and improve resource utilization. Several studies focus on modeling and scheduling mechanisms to improve spot instance reliability \cite{modeling_HPDC20, spot_yang_www22}. For instance, Snape \cite{Snape} employs reinforcement learning to dynamically optimize hybrid deployments of spot and on-demand instances. 
Additionally, many studies emphasize application-level adaptations, particularly for computationally intensive workloads \cite{DeepSpotCloud, Varuna, miao2024spotserve}. DeepSpotCloud \cite{DeepSpotCloud} and Varuna \cite{Varuna} address the use of spot instances for DL training, with continuous checkpointing and redundant computation to cope with frequent preemption. SpotServe \cite{miao2024spotserve} specifically targets distributed LLM inference using preemptible instances. 

However, these works overlook two key considerations. First, prior modeling efforts mainly focus on CPU-centric workloads, neglecting GPU constraints, especially the static quotas set by cloud providers that limit spot allocations. Second, application-focused solutions take a user-centric approach, optimizing individual tasks instead of addressing resource efficiency across the cluster. Our work addresses these gaps by introducing a proactive cluster management framework from the administrator’s perspective.

\subsection{GPU Scheduling in Large-Scale Clusters}

Existing research on GPU scheduling can be divided into two categories. The first category focuses on minimizing preemption overhead through runtime optimization \cite{Lightweight, PipeSwitch}. For instance, PipeSwitch \cite{PipeSwitch} employs lightweight context-switching mechanisms to accelerate task migration during preemption. 
The second category prioritizes deadline-aware resource allocation, exemplified by SHEPHERD \cite{SHEPHERD} and Chronus \cite{Chronus}, which dynamically balance preemption risks with job scheduling to meet completion deadlines. While Lyra \cite{Lyra} reduces preemption costs via heuristic scheduling, it lacks proactive demand analysis.
Checkpointing strategies further enhance fault tolerance in such systems. CheckFreq \cite{CheckFreq} optimizes checkpoint intervals to balance computational overhead and recovery efficiency. However, these approaches primarily address reactive fault recovery rather than integrating checkpointing with proactive scheduling.

\section{Conclusion}

\label{sec:conclusion}
 
LLMs pose new challenges in the management of GPUs in large-scale clusters. An extensive analysis of over 10,000 GPUs identified critical limitations in existing schedulers, particularly regarding high eviction rates and long queuing times. To tackle these issues, we propose GFS, a novel preemptive scheduling framework designed to optimize SLO compliance for HP tasks while minimizing preemption of spot tasks. GFS has been validated through both real-world production deployments and trace-driven simulations. In the future, we will focus on optimizing the utilization of computing units (e.g., SMs and tensor cores) in GPU, to further enhance overall efficiency in large-scale clusters.


\section{Acknowledgement}

We thank our shepherd Mohammad Shahrad and anonymous reviewers for their valuable comments. This work was supported by Alibaba Tech infra and Reliability Engineering (TRE) in Alibaba Group through Alibaba Innovative Research Program and Alibaba Research Intern Program. This work was also supported by the ``Pioneer'' R\&D Program of Zhejiang (2025C01001).


\bibliographystyle{plain}
\balance
\bibliography{ref}


\end{document}